\documentclass[11pt,oneside,english]{amsart}
\usepackage{geometry}
\usepackage{amsmath,amsfonts,dsfont}
\usepackage{appendix}
\usepackage{amssymb}
\usepackage{tikz}
\usepackage[T1]{fontenc}
\usepackage[latin9]{inputenc}
\usepackage{color}
\usepackage{latexsym}
\usepackage{amsbsy}
\usepackage{amstext}
\usepackage{amsmath}
\usepackage{amssymb}
\usepackage{tikz}
\usepackage{amsthm}
\usepackage{amstext}
\usepackage{amssymb}
\usepackage{graphicx}
\usepackage{epstopdf}
\usepackage{esint}
\usepackage{babel}

\usepackage{times}
\usepackage{color}
\usepackage{fancyhdr}
\AtBeginDocument{

}
\begin{document}
\global\long\def\bK{\mathbb{K}}
 \global\long\def\bC{\mathbb{C}}
 \global\long\def\bR{\mathbb{R}}
 \global\long\def\bZ{\mathbb{Z}}
 \global\long\def\bN{\mathbb{N}}
 \global\long\def\bQ{\mathbb{Q}}
 \global\long\def\bH{\mathbb{H}}

 \global\long\def\half{\frac{1}{2}}
 \global\long\def\ii{\mathrm{i}}

 \global\long\def\bdry{\partial}
 \global\long\def\cl#1{\overline{#1}}

\global\long\def\PR{\mathsf{P}}
 \global\long\def\EX{\mathsf{E}}
 \global\long\def\sU{\mathcal{U}}

\global\long\def\GL{\mathrm{GL}}
 \global\long\def\SL{\mathrm{SL}}
 \global\long\def\gl{\mathfrak{gl}}
 \global\long\def\sl{\mathfrak{sl}}
 \global\long\def\TL{\mathrm{TL}}

\global\long\def\Hom{\mathrm{Hom}}
 \global\long\def\End{\mathrm{End}}
 \global\long\def\Aut{\mathrm{Aut}}
 \global\long\def\Rad{\mathrm{Rad}}
 \global\long\def\Ext{\mathrm{Ext}}

\global\long\def\Kern{\mathrm{Ker}}
 \global\long\def\Imag{\mathrm{Im}}

\global\long\def\Alt{\mathrm{Alt}}

\global\long\def\dmn{\mathrm{dim}}
 \global\long\def\spn{\mathrm{span}}
 \global\long\def\tens{\otimes}
 \global\long\def\Mat{\mathrm{Mat}}
 \global\long\def\unitmat{\mathbb{I}}
 \global\long\def\id{\mathrm{id}}
 \global\long\def\diag{\mathrm{diag}}
 \global\long\def\unit{\mathbf{1}}

\global\long\def\set#1{\left\{  #1\right\}  }
 \global\long\def\sgn{\mathrm{sgn} }

\global\long\def\re{\Re\mathfrak{e}}
 \global\long\def\im{\Im\mathfrak{m}}
 \global\long\def\arg{\mathrm{arg}}
 \global\long\def\isom{\cong}

\global\long\def\op{\mathrm{op}}
 \global\long\def\cop{\mathrm{cop}}

\global\long\def\eps{\varepsilon}
 \global\long\def\const{\mathrm{const.}}

\global\long\def\binomial#1#2{{#1  \choose #2}}

\global\long\def\Hspace{\mathcal{F}}
 \global\long\def\Hinner#1#2{\left\langle #1,\,#2\right\rangle _{\Hspace}}

\global\long\def\spin{\sigma}
 \global\long\def\spinconf{\boldsymbol{\sigma}}
 \global\long\def\energy{\varepsilon}
 \global\long\def\contourset{\mathcal{C}}

\global\long\def\rect{R}
 \global\long\def\rectTop{\mathrm{top}}
 \global\long\def\rectBot{\mathrm{bot}}

\global\long\def\NE{\mathrm{NE}}
 \global\long\def\NW{\mathrm{NW}}
 \global\long\def\SW{\mathrm{SW}}
 \global\long\def\SE{\mathrm{SE}}

\global\long\def\iinterval#1{\llbracket\,\!#1\,\!\rrbracket}
 \global\long\def\hiinterval#1{\llbracket\,\!#1\,\!\rrbracket^{*}}

\global\long\def\Srow{\mathrm{Srow}}
 \global\long\def\Cliff{\mathrm{Cliff}}
 \global\long\def\CliffGen{\mathrm{Cliff}^{(1)}}
 \global\long\def\RH{\mathrm{RH}}
 \global\long\def\sP{\mathcal{P}}
\global\long\def\SLE{\mathrm{SLE}}
 \global\long\def\SLEk{\mathrm{SLE}_{\kappa}}

\global\long\def\iinterval#1{\llbracket\,\!#1\,\!\rrbracket}
 \global\long\def\hiinterval#1{\llbracket\,\!#1\,\!\rrbracket^{*}}

 \global\long\def\Srow{\mathrm{Srow}}
 \global\long\def\Cliff{\mathrm{Cliff}}
 \global\long\def\CliffGen{\mathrm{Cliff}^{(1)}}
 \global\long\def\RH{\mathrm{RH}}
\global\long\def\sP{\mathcal{P}}

\numberwithin{equation}{section}

\title{New Phase Transitions in Chern-Simons Matter Theory}
\author{Ali Zahabi}
\begin{abstract}
Applying the machinery of random matrix theory and Toeplitz determinants we study the level $k$, $U(N)$ Chern-Simons theory coupled with fundamental matter on $S^2\times S^1$ at finite temperature $T$. This theory admits a discrete matrix integral representation, i.e. a unitary discrete matrix model of two-dimensional Yang-Mills theory. In this study, the effective partition function and phase structure of the Chern-Simons matter theory, in a special case with an effective potential namely the Gross-Witten-Wadia potential, are investigated.
We obtain an exact expression for the partition function of the Chern-Simons matter theory as a function of $k,N,T,$ for finite values and in the asymptotic regime. In the Gross-Witten-Wadia case, we show that ratio of the Chern-Simons matter partition function and the continuous two-dimensional Yang-Mills partition function, in the asymptotic regime, is the Tracy-Widom distribution. Consequently, using the explicit results for free energy of the theory, new second-order and third-order phase transitions are observed. Depending on the phase, in the asymptotic regime, Chern-Simons matter theory is represented either by a continuous or discrete two-dimensional Yang-Mills theory, separated by a third-order domain wall.
\end{abstract}
\maketitle

\section{Introduction}
Study of the phase structure of $(p+1)-$dimensional Yang-Mills (YM) theory coupled to matter on $S^p$ sphere and especially the confinement/deconfinement transition in large $N$ gauge theories on sphere sheds light on various phenomena in gravitational phase transition such as black hole nucleation, via the AdS/CFT correspondence. In this direction, the phase structure of Chern-Simons (CS) theory coupled to matter and its gravitational dual, Vasiliev higher spin gravity, \cite{Va}, have been studied recently, \cite{Ch-Mi} and \cite{Gi}.
In a recent work \cite{Ta}, it has been argued that level $k$, $U(N)$ CS theory coupled to fundamental matter on $S^2\times S^1$ at finite temperature $T$, can be explained by a discrete unitary matrix model, and its phase structure, the eigenvalue density in different phases, has been studied in the limit of large $k,N,T$, with fixed parameters $\lambda=N/k, \zeta=T/N$, by using steepest descent method. This study is based on the reduction of the Chern-Simons matter (CSM) theory first to the CS theory with an effective potential and second to a discrete version of two-dimensional YM theory and its representation, one matrix model. In a special case with Gross-Witten-Wadia (GWW) effective potential, \cite{Gr-Wi} and \cite{Wa}, in addition to well-known lower-gap phase, an upper-gap phase is observed, \cite{Ta} and \cite{Tak}, in which the upper bound of the eigenvalue density is saturated.

In another discipline of research, discrete Toeplitz determinants are studied in \cite{Ba-Li}, and by using the continuous orthogonal polynomials their precise forms are obtained in terms of the continuous Toeplitz determinants and the Fredholm determinant. A key fact to understand the application of the Toeplitz determinant in the CSM theory is the Heine-Sezg\"{o} identity which relates the matrix model integral and the Toeplitz determinant, see appendix A.

Using the relation between Toeplitz determinants and matrix integrals, and the above results for discrete matrix model representation of the CSM theory, in this paper, the techniques and methods from discrete Toeplitz determinant is employed to study the CSM theory. The matrix integral representations of the YM theory and CSM theory provide the application of well-developed techniques of random matrix theory and specially the Toeplitz determinant.

In our setting, the partition functions of YM theory and CSM theory with an arbitrary potential $V(T)$ can be written as Toeplitz determinant and discrete Toeplitz determinant with respect to the probabilistic weight $f(T)=e^{-V(T)}$, respectively,
\begin{equation}\label{eq: gauge PF and TD}
\mathcal{Z}_{YM}:=\mathcal{Z}_{N}(V)=\mathrm{D}_N(f), \hspace {.5cm} \mathcal{Z}_{CSM}:=\mathcal{Z}_{N}^k(V)=\mathrm{D}_N^k(f).
\end{equation}

In this study, we move towards an exact determination of the thermal partition function of CSM theory. In fact, by adopting recent results, \cite{Ba-Li}, about the discrete and continuous Toeplitz determinants, a new explicit expression for the partition function of CSM theory in terms of the partition function of YM theory, as a function of $k,N,T$ at the finite values and in the asymptotic regime, is obtained. Especially, a CSM theory with GWW effective potential is studied in detail and analytic forms of the effective partition function and free energy of CSM theory are obtained. Furthermore, the sub-leading corrections in the asymptotic results are computed and the result is interpreted as the energy of the upper bound. These results lead to a phase transition in the asymptotic regime of the discrete matrix model, related to the phase transition between upper-bound phase and no-upper bound phase.

Although, the effective GWW potential is not a microscopic potential and it does not exactly characterize the CSM theory, but the CSM model with GWW potential shares some features with the real CS theory such as the minimum of any effective potential $v(U)$ (as a function of holonomy $U$) is always at $U=I$ and the depth of the potential increases as a function of $\zeta$.

In brief, the phase structure of the model in the asymptotic limit of large $N,T,k$ is studied via the asymptotic behavior of the ratio of partition functions which is obtained as the Tracy-Widom distribution. Then, from analytic expression for the free energy, obtained from the expansion of Tracy-Widom distribution, new second-order and third-order phase transitions in this model are determined. Moreover, relation between the obtained result in this paper and other recent results is explored.

The phases of CSM theory, can be discussed from the viewpoint of the existence of the continuum limit. In fact, the continuum limit does not exist in whole moduli space of the parameters and in fact, $N,T,k$ should satisfy certain relations in order to determine the continuum limit.

This paper is organized as follows:
In section two, the CSM theory and the discrete matrix integral representation of the partition function are explained in brief. In section three, the main results of this study are expressed. First, the partition function and free energy of the CSM theory with GWW potential are explicitly obtained and subsequently the phase structure of the model with GWW potential is studied. In section four, possible interpretations of the obtained results are discussed and directions for further studies are introduced. Finally, there are two appendices, where necessary materials from the random matrix theory such as orthogonal polynomials, Toeplitz determinant and Riemann-Hilbert problem (RHP) and their inter-relations are introduced and well-known results in two-dimensional YM theory with GWW potential in terms of Toeplitz determinant and RHP problem, and some results about the partition function and phase structure in the CSM theory are summarized. Finally, some necessary definitions and results about Tracy-Widom distribution are collected.
\section{Chern-Simons matter theory}
In this section, the level $k$, $U(N)$ CSM theory coupled to fundamental matter at temperature $T$ on $S^2\times S^1$ is described. Especially, the CS theory coupled with the matter in fundamental representation on $S^2\times S^1$, in $T,N,k\rightarrow \infty$ limit with the fixed 't Hooft parameter, $\lambda=N/k$, and $\zeta=T/N$, will be studied. In this part, we explain two relevant and necessary results from the literature; I) the action of pure CS theory coupled with fundamental matter in the 't Hooft limit and at high temperature, up to leading order in $N$, can be expressed as the sum of the action of pure CS theory and an effective action, i.e. effective potential which is a local function of holonomy matrix $U(x)$ with eigenvalues $e^{\ii\alpha}$ around $S^1$ with $x\in S^2$. II) The partition function of the CSM theory with effective potential can be written as a discrete version of the matrix integral representation of the $2d$ YM theory with the effective potential. These observations are explained in this section and in the Appendix A, not only for the article to be self contained but also because of their importance and usefulness in understanding the physics and mathematics of the problem studied in this article, since our new results are based on these observations.

\subsection{CSM theory and discrete YM theory}
CS theory is a three-dimensional topological field theory of the Schwarz type \cite{Sc}, with no metric dependence in the action.
Consider a three-dimensional manifold $M$ with a principal $G-$bundle with a connection $A$, then the CS action is defined by
\begin{equation}\label{eq: CS action}
\mathcal{S}_{CS}(A)=\frac {k}{4\pi}\int_M Tr (A\wedge dA+\frac {2}{3}A\wedge A\wedge A),
\end{equation}
where $k$ is called level and it is an integer in the quantized theory. The above action functional defines a three-dimensional topological field theory, i.e. a field theory with observables, explicitly independent of the metric on $M$.
The partition function of the theory is given by a path integral,
\begin{equation}\label{eq: CS partition function}
\mathcal{Z}_{CS}(A)=\int [\mathcal{D}A]\ e^{\ii\mathcal{S}_{CS}},
\end{equation}
and the observables of the theory are defined via the correlation functions of the Wilson loops, $\mathcal{W}_{C_i, R_i}(A)=Tr\left[\mathcal{P}e^{\ii\oint_{C_i}A}\right]$ in $R_i$ representation with closed loops $C_i$, as,
\begin{equation}\label{eq: CS observable}
\left<\prod_{i=1}^n \mathcal{W}_{C_i, R_i}(A)\right>=\int [\mathcal{D}A]\ e^{\ii\mathcal{S}_{CS}}\left(\prod_{i=1}^n \mathcal{W}_{C_i, R_i}(A)\right).
\end{equation}
In this study, we consider $G=U(N)$. The partition function and correlation functions of the theory define global topological invariants of the manifold as well as other topological invariants associated to the manifold such as knot invariants, Jones polynomials, etc. \cite{Wi}.
In this study, we are interested in CS theory on curved background $S^2\times S^1$. Such curved backgrounds provide enough symmetry for explicit computation of the partition function and correlation functions. For example, the partition function of CS theory on Seifert manifolds such as $S^3$ admits a matrix model representation and thus analytic closed expressions for the partition function have been obtained in \cite{Ma1}. For a general review on this and other different aspects of CS theory, see \cite{Ma2}.

The gauge theory in this study is a level $k$, $U(N)$ CS theory at finite temperature and finite $N$, as well as its high temperature and large $N$ limit, coupled to fundamental matter fields. The behavior of the CSM theories and their partition functions and free energies as functions of $\zeta$ and $\lambda$ has been a subject of intense studies during past years. In the limit $\lambda\rightarrow 0$, the behavior of CSM theory and its thermal free energy has been studied in \cite{Sh-Yi}, \cite{Gi}, \cite{Ch-Mi}. Fundamental matter CSM theories at finite 't Hooft coupling $\lambda$ and at large $N$ have been studied in \cite{Gi}, \cite{Ch-Mi}, \cite{Ah-Gu-Ya1}, \cite{Ah-Gu-Ya2}, \cite{Ma-Zh1}, \cite{Ma-Zh2}, \cite{Ba-He}, \cite{Ja}, \cite {Yo}, \cite{Gu}, \cite{Ah}. The phase structure and thermal free energy of some examples of such theories have been studied in \cite{Ta}, \cite{Tak}.

The partition function of the CSM theories contain the CS action and a matter action which shows the interaction with the matter fields,
\begin{equation}\label{eq: CSM partition function}
\mathcal{Z}_{CSM}(A,\mu)=\int [\mathcal{D}A] [\mathcal{D}\mu] \ e^{\ii\mathcal{S}_{CS}-\mathcal{S}_{\mu}},
\end{equation}
where $\mu$ represents the matter fields and $\mathcal{S}_{\mu}$ is the corresponding action. The matter content of the CSM theory consists of vector massive fields in the fundamental representation such as fermionic and bosonic fields minimally coupled to CS fields, for further details see references in the previous paragraph. In this study we will not discuss any computations containing explicit forms of the matter fields, thus we will keep to our general form of matter action and we will not be more explicit about it. As we will see in the following, in the high temperature limit, the matter fields have masses of order the temperature and the effective action only depends on the holonomy field.

The goal of this part is to compute the partition function of CSM theory. In order to do this computation, roughly speaking, we have to sum all the vacuum graphs of the massless and massive fields and thus the partition function of CSM theory is obtained by summing up all the separate contributions from the vacuum graphs of massive and massless fields including the holonomy field. In brief, we first evaluate the path integral on the matter fields which gives us the partition function as a path integral over massless fields with a CS action and an effective action for the holonomy field. The effective action is directly obtained from path integration over the massive fields. More precisely, after integration over the massive fields, a local effective action, $\mathcal{S}_{eff}(U)$, for the $2d$ unitary matrix valued holonomy $U(x)$ around $S^1$ with $x\in S^2$ is obtained and the CSM partition function takes the following form,
\begin{equation}
\mathcal{Z}_{CSM}(A)=\int [\mathcal{D}A] \ e^{\ii\mathcal{S}_{CS}-\mathcal{S}_{eff}(U)}.
\end{equation}

Here, we briefly discuss the behavior of the effective action such as its series expansion and its leading term in the large $N$ limit which is a local function of holonomy field.  The key fact in evaluation of the effective action and CSM partition function is the following observation in \cite{Ah}. In order to obtain the effective action, we integrated out the vacuum graphs of the matter fields and since these matter fields develop large thermal masses, thus $\mathcal{S}_{eff}(U)$ can be expanded as the following local series,
\begin{equation}
\mathcal{S}_{eff}= \int d^2x (T v(U (x))+v_1(U)Tr D_iU D^iU+...,
\end{equation}
where $v(U(x))$ and $v_1(U(x))$ are local effective potentials which their actual forms depend on the matter content and CS couplings. In the large $N$ limit, the explicit computation of the effective potentials as a function of $\lambda$, for any given vector matter CS theory is feasible. For example, for CSM theory with $U=I$, and minimally coupled fundamental fermions to the CS fields, the effective potential computed by using large $N$ techniques in the light-cone gauge, \cite{Gi}. This computation has been generalized to other theories and different holonomy matrices, \cite{Ja}, \cite{Ah}, \cite{Ta} and references therein. It has been argued in \cite{Ah}, that the effective potentials are of order $N$, and since $T=\zeta N$, the first term in the expansion of the effective potential is of order $N^2$ and the second one is of order $N$ and the rest are further suppressed at large $N$. Therefore, at large $N$ with $T\sim N$, up to leading order in $N$, the entire effect of matter loops in vacuum graphs on the dynamics of the CS theory is represented by the high temperature effective action of the following form,
\begin{equation}
\mathcal{S}_{eff}= T\int d^2x\sqrt {g}v (U (x)).
\end{equation}
And finally the thermal partition function of the CSM theory becomes
\begin{equation}\label{eq: CSM effective partition function}
\mathcal{Z}_{CSM}=\int[\mathcal{D}A]e^{\ii\frac {k}{4\pi}\int_M Tr (AdA+\frac {2}{3}A^3)-T\int d^2x\sqrt {g}v (U (x))}=<e^{-TV_2 v(U)}>_{N,k},
\end{equation}
where $V_2$ is the $2d$ volume of $S^2$ and since the pure CS is topological, the expectation value of the exponential of the integral of the effective potential over $S^2$ is independent of $x$. Therefore, the CSM partition function can be written as a topological observable of the pure CS theory, i.e. an expected value of linear combination of the Wilson loops in pure CS theory. For more convenience assume $V_2=1$.

An explicit, matrix integral representation of the CSM partition function (\ref{eq: CSM effective partition function}) can be obtained by using the path integral techniques, \cite{Ta}, \cite{Bl-Th}, see Appendix A. Roughly speaking, after gauge fixing and evaluating the integral of the CS action over $S^2$, the path integral over the space of gauge fields reduces to a matrix integral over the eigenvalues of the holonomy matrix. This matrix integral representation of $\mathcal{Z}_{CSM}$ is a discrete version of the matrix integral representation of the two-dimensional YM theory partition function, Eq. (\ref{eq: Matrix integral for Yang Mills}) with $V_{YM}=N\zeta v(U)$.
As a result of this computation, partition function of CSM theory, Eq. (\ref {eq: CSM effective partition function}), is given by a discrete unitary matrix integral and equivalently, by using Heine-Szeg\"{o} identity, as a discrete Toeplitz determinant,
\begin{eqnarray}\label{eq: effective CSM partition function}
\mathcal{Z}_{CSM}&=& \int [\mathcal{D}A] \ e^{\ii \mathcal{S}_{CS}- T\int d^2x\sqrt {g}v (U (x))}\nonumber\\
&=&\prod_{i=1}^N\sum_{n_i=-\infty}^\infty [\prod_{l< i} \left(2\sin(\frac {\alpha_l(\overrightarrow {n})-\alpha_i(\overrightarrow{n})}{2})\right)^2e^{-N\zeta v(U)}]\nonumber\\
&=&\det\left[\frac {1}{k}\sum_{z\in d_1}z^{-j+l} f (z)\right]_{j, l=0}^{N-1}\nonumber\\
&=&\mathrm{D}_N^k (f,d_1),
\end{eqnarray}
where $n_i$'s are integer and $\alpha_i( \overrightarrow{n})=\frac{2\pi n_i}{k}$ are the eigenvalues of the holonomy matrix, (distributed on the unit circle with distance $2\pi/k$ between two consecutive eigenvalues), and $z=e^{\ii\alpha}$  and $d_1=\{z\in \mathds{C}| z^k=1\}$ is a finite discrete subset of a unit circle $S^1$ with size $|d_1|=k$ and $f=e^{-N\zeta v(U)}$ is the weight function.

A toy model of the CSM theory, Eq. (\ref{eq: effective CSM partition function}), with interesting features is the CS theory with the GWW potential, $v (U)=-\frac{1}{2}Tr(U+U^\dagger)$, or $v(\alpha)=-\sum_{i=1}\cos \alpha_i$.
In the following section, based on the above Toeplitz determinant formulas, exact analytic formulas for the free energy of the CSM theory with GWW potential are obtained. Furthermore, using the obtained explicit results for the free energy, the phase transitions and their orders between different phases of the theory are studied.
\section{New results in Chern-Simons matter theory}
In this part, the mathematical results in Toeplitz determinants have been adopted and translated in the language of gauge theory and their deep implications and applications in the study of gauge theory, e.g. in determining the free energy of the theory and revealing the phase structure have been studied. As we will see, this new method in study of gauge theory is not only able to reproduce a very recent important result about the phase structure of CSM, produced by other plausible methods in \cite{Ta} and \cite{Tak}, but also provides explicit expressions for the free energy of the gauge theory which contains important new results for the order of phase transitions in gauge theory.

In this section, a new explicit results for the free energy and new phase transitions in the CSM theory with GWW potential are obtained. In order to obtain the final result, in the first part, the partition function of the CSM theory with an arbitrary potential is obtained via a recent result in the context of the discrete Toeplitz determinants. In the second part, the CSM theory with GWW potential is studied and explicit asymptotic results, in the limit of large parameters, for the partition function is obtained. In the third part, a careful study of the obtained results leads to an explicit expression for the free energy of the CSM theory with GWW potential which consequently reveals new phase transitions in this case. In the fourth part, some consistency checks of the obtained results as well as some explicit computations for the free energy in different limits are performed.
\subsection{CSM theory partition function}
It has been recently shown that the analysis of discrete Toeplitz determinants can be done by means of continuous orthogonal polynomials associated to continuous weight function, \cite{Ba-Li}. The main result in that study is the explicit relation between the discrete and continuous versions of the Toeplitz determinant for a given weight function in terms of a Fredholm determinant. In this study, we investigate the implications of those results for gauge theory. In other words, via the relation between the Toeplitz determinants and partition functions, see appendix A, the relation between the discrete and continuous Toeplitz determinants is used to connect the partition functions of CSM theory (discrete YM theory) and continuous YM theory. The physical meanings and implications of this connection for phase structure of the CSM theory are studied.

In order to make the connection between Toeplitz determinants and CSM gauge theory precise, we studied the relations between the parameters in both subjects. In both subjects, there are parameters for indicating the features of the systems and these parameters are closely related by their meanings in each subject. For example, the rank of the gauge group in CSM theory is equivalent to the size of the Toeplitz determinant and the level of the CSM theory is equivalent to a parameter in discrete Toeplitz determinant that shows the discreteness. The temperature of CSM theory is an extra parameter in the weight function of the Toeplitz determinant. In summary, there is a one to one correspondence between the parameters in CSM theory and Toeplitz determinants which can be easily observed by comparing two subject.

After adopting the results in discrete Toeplitz determinants for the gauge theory in this study, the following explicit formula for the partition function is obtained from Eq. (\ref{eq: PaFu and ToDe}) and Eq. (\ref{ratio Fredholm}) in Appendix A. In our setting, $z=e^{\ii \alpha}$ and domain $d_1= \{z\in \mathds{C}| z^k=1\}$ is a finite discrete subset of a unit circle $S^1$ with size $|d_1|=k$ (level of CS theory plays the role of discreteness in Toeplitz determinant) and $f(z)=e^{-V(z)}$ is a weight function, then, Eq. (\ref{eq: PaFu and ToDe}) and Eq. (\ref{ratio Fredholm}) imply
\begin{equation}\label{CSM partition function Fredholm}
\frac{\mathcal{Z}_{CSM}}{\mathcal{Z}_{YM}}=\frac{\mathrm{D}_N^k (f, d_1)}{\mathrm{D}_N (f)}=\det {\left(1+K\right)},
\end{equation}
where $K$ is the integral operator with a kernel,
\begin{equation}\label{eq: simplified kernel}
K (z, w)=z^{-N}\frac {p_N (z)p_N^*(w)- p_N^* (z)p_N(w)}{1-z^{-1}w}\sqrt {v_k (z)v_k (w)e^{-V(z)}e^{-V(w)}},
\end{equation}
and $p_N(z)$ is an orthogonal polynomial with respect to weight function $e^{-V(z)}$, (see Appendix (A.2)) and $p^*_N(z):=z^N \overline{p_N(\bar{z}^{-1})}$ (on the circle $S^1$, $\bar{z}^{-1}=z$), and by using $\gamma(z)=z^k-1$ in our domain, discrete function $v_k(z)$ is obtained from Eq. (\ref{general discrete function}) as,
\begin{equation}\label{discrete function}
v_k(z)=
\begin{cases}
-\frac{z^k}{1-z^k}&  \ z\in S^1_{\textit{in}}\\
\frac{z^{-k}}{1-z^{-k}}& \ z\in S^1_{\textit{out}}
\end{cases}.
\end{equation}
Finally, the Fredholm determinant for the integral operator $K$ is formally defined by,
\begin{equation}
  \det(1+K)=1+\sum_{n=1}^{\infty}\frac{(-1)^n}{n!}\int\int...\int\det\left(K(z_i, z_j)\right)_{i,j=1}^{n}dz_1...dz_n.
\end{equation}

Up to a normalization, the free energy of CSM theory is obtained by taking the logarithm of Eq. (\ref{CSM partition function Fredholm}),
\begin{equation}\label{eq: general free energy of CSM}
\mathcal{F}_{CSM}= \mathcal{F}_{YM}+\log{\det {\left(1+K\right)}}.
\end{equation}
For a trace class operator $K$, we have,
\begin{equation}
\mathcal{F}_{CSM}= \mathcal{F}_{YM}+Tr(\log{(1+K)})=\mathcal{F}_{YM}+\sum_{n=1}^{\infty}(-1)^{n+1}\frac{Tr K^n}{n}.
\end{equation}
The above formal expression (\ref{eq: general free energy of CSM}) is a new exact result for the free energy of the CSM theory at any values of $N,k,T$. In principle, it contains more information about CSM theory in comparison to the existing results in the literature which only discuss the free energy in the large $N,k,T$ limit. In fact, its explicit evaluation reduces to evaluation of the Fredholm determinant with a kernel associated to the effective potential of the CSM theory. Moreover, Eq. (\ref{eq: general free energy of CSM}) basically contains explicit information about the phase structure of the CSM with any potential described in Eq. (\ref{eq: effective CSM partition function}). Roughly speaking, the kernel $K$ which depends on the weight function $f$ contains information about the free energy and it determines the phase structure of the theory.

In fact, as it will be clear in the following section in the example of CSM theory with GWW potential, by using this new method we can explicitly obtain the free energy and the phase structure of the model. And since the CSM model with GWW potential shares some features with the real CS theory we expect that CSM theory with GWW potential has the main qualitative features of the phase structure of realistic CSM theories with other different potentials. The CSM theory with any potential and its phase structure can be exactly studied by using the explicit form of the free energy given by Eq. (\ref{eq: general free energy of CSM}). However, in this study we focus on the toy model with GWW potential as the first step towards CSM theory with other complicated potentials.

In gauge theories, we are interested in the limit of large parameters, in which the ratio of the partition functions can be studied via the asymptotic analysis of the Fredholm determinant. This analysis is mainly based on the asymptotic analysis of the orthogonal polynomials, \cite{Sz}, \cite{Ma-Mc}. In this study, instead of repeating the standard asymptotic analysis of the Fredholm determinant we rather prefer to focus on the physical aspects of the obtained results from these analysis. As the simplest example of such analysis, we consider a case of a positive constant weight function, $f_c=const.$ In this case, the asymptotics of orthogonal polynomial as $N\rightarrow\infty$, \cite{Sz}, is
\begin{equation}
p_N(z)=
\begin{cases}
z^N\mathcal{O}(e^{-C N}) & \  |z|\geq 1+\epsilon\\
\mathcal{O}(e^{-C N})& \ |z|\leq 1-\epsilon
\end{cases},
\end{equation}
where $C$ is a constant, and by using Eq. (\ref{discrete function}), the discrete function $v_k(z)$ for large $k$ becomes
\begin{equation}
v_k(z)\leq
\begin{cases}
2(1-\epsilon)^k&  \ |z|=1-\epsilon\\
2(1+\epsilon)^{-k}& \ |z|=1+\epsilon
\end{cases}.
\end{equation}
Above asymptotic formulas for orthogonal polynomials and discrete function, imply the following asymptotic results (in the limit $N,k\rightarrow\infty$) for the ratio of partition functions, Eq. (\ref{CSM partition function Fredholm}), with constant potential,
\begin{equation}\label{simple ratio}
\lim_{k-N\rightarrow \infty, N\rightarrow \infty}\frac{\mathcal{Z}_N^k (f_c, d_1)}{\mathcal{Z}_N (f_c)}=1+\mathcal{O}(e^{-c (k+N)}),
\end{equation}
where $c$ is a positive constant and we have used Eq. (\ref{eq: gauge PF and TD}). The analysis details of the proof of the above result can be found in \cite{Ba-Li}. We are interested in the physical meaning of the above result. In fact, Eq. (\ref{simple ratio}) is expected, because, when the discrete structure vanishes in the limit $k\rightarrow \infty$, then, in the leading order, the partition function of the continuum YM theory is the continuum limit of partition function of the discrete YM theory. In other words, for a constant potential, we expect free energy be a smooth function in the asymptotic limit. The continuum limit is smooth in the sense that we do not expect any discontinuity in the behavior of CSM theory that leads to a phase transition in the continuum limit. However, this soft continuum limit is not always the case and as it will be studied in this paper, the asymptotic limit of the ratio (\ref{CSM partition function Fredholm}) depends on the weight function in a nontrivial way and the continuum limit is only obtained in a special region of the moduli space of the parameters.
\subsection{Partition function and phase structure of CSM theory with GWW potential}
In this part, we explicitly demonstrate the usefulness of Eq. (\ref{CSM partition function Fredholm}) in gauge theory and its ability to produce physical results in CSM theory. In an important example, the partition function of CSM theory with GWW potential is obtained explicitly in terms of the partition function of the GWW model, (see Appendix A.1.2), and by careful analysis of the asymptotic regime, new phase transitions are traced.

In the finite regime, the following equation, obtained from Eq. (\ref{CSM partition function Fredholm}), determines the CSM partition function with GWW potential,
\begin{equation}\label{eq: CSM GW PF}
\mathcal{Z}^{(GWW)}_{CSM}=\mathcal{Z}^{(GWW)}_{YM} \det{(1+K_{GWW}(z, w))},
\end{equation}
where $K_{GWW}(z, w)$ in principle can be obtained from Eq. (\ref{eq: simplified kernel}) by putting $V_{GWW}(z)=-\frac{T}{2}(z+z^{-1})$ and using appropriate $p_N(z)$ obtained from Eq. (\ref{orthogonal polynomials}) with $V_{GWW}$, see \cite{Ba-Li}, and $v(z)$ from Eq. (\ref{discrete function}).

In the limit $N, T, k\rightarrow \infty$, asymptotic results for the ratio, Eq. (\ref{CSM partition function Fredholm}), in the case of GWW potential can be obtained either from a direct asymptotic analysis of the Fredholm determinant, similar to the analysis for the case of the constant potential in the previous part, or by using the adopted version of the results for Toeplitz determinants in the context of Brownian motion, see Appendix A.1.1. In this study, we will not repeat any details of such analysis in the Brownian motion, instead we formally use the results in Brownian motion as some mathematical results and we further explore their implications for gauge theory. As a mathematical result, in domain $d_s= \{z\in \mathbb{C}|\ z^k=s\}$, for $f_{GWW}=e^{-V_{GWW}}$ by putting together Eq. (\ref{eq: conditional probability}) and Eq. (\ref{eq: asymptotics of conditional probability}), we obtain,
\begin{equation}\label{eq: GW PF TW}
\lim_{N,T,k\rightarrow\infty} \oint_{|s|=1}\frac{\mathcal{Z}_N^k(f_{GWW},d_s)}{\mathcal{Z}_N(f_{GWW})}\frac{ds}{2\pi\ii s}= F(\frac{k-\mu}{\sigma}),
\end{equation}
where $F$ is the Tracy-Widom distribution (see appendix B) and functions $\mu$ and $\sigma$ are defined by
\begin{equation}
\mu :=
\begin{cases}
N+T&  \ N\geq T\\
2\sqrt {NT}& \ N<T
\end{cases},\hspace{.5 cm}
\sigma :=
\begin{cases}
 2^{-\frac{1}{3}} T^{\frac {1}{3}}& \  N\geq T\\
 2^{-\frac{2}{3}}T^{\frac {1}{3}}(\sqrt {\frac {N}{T}}+\sqrt {\frac {T}{N}})^{\frac {1}{3}}& \  N<T
\end{cases}.
\end{equation}

For $s=1$, Eq. (\ref{eq: GW PF TW}), implies the following new and useful result in gauge theory which is an explicit relation between the partition functions of CSM theory and YM theory with GWW potential,
\begin{equation}\label{eq: ratio for GW}
\lim_{N,T,k\rightarrow \infty}\frac{\mathcal{Z}^{(GWW)}_{CSM}}{\mathcal{Z}^{(GWW)}_{YM}}=F(\frac{k-\mu}{\sigma}).
\end{equation}
As a remark, by using an expression for the Tracy-Widom distribution in terms of the kernel of the Airy function, $F=\det(1+K_{Air})$, (see Appendix B), Eqs. (\ref{eq: CSM GW PF}) and (\ref{eq: ratio for GW}), imply that $\lim_{N,T,k\rightarrow \infty}K_{GWW}(z,w)=K_{Ai}(z, w)$.

In the following, we extract the physical meaning behind the simplicity and elegancy of the above mathematical result, Eq. (\ref{eq: ratio for GW}). In the next step, an asymptotic formula for the ratio of the partition functions, Eq. (\ref{eq: ratio for GW}), is obtained via the asymptotic formula for the Tracy-Widom distribution. Let us denote $x=\frac{k-\mu}{\sigma}$, then it is easy to see that the asymptotic limit $k,N,T\rightarrow\infty$, leads to $x\rightarrow\pm\infty$ for $k>\mu$ and $k<\mu$, respectively. The asymptotic formula for the Tracy-Widom distribution, Eq. (\ref{eq: asymp of Tracy Widom}) in Appendix B, is
\begin{equation}\label{eq: leading asymp of F}
F(x)=
\begin{cases}
1-\mathcal{O}(e^{-x^{3/2}})& \  x\rightarrow \infty\\
\mathcal{O}(e^{-|x|^3})& \  x\rightarrow -\infty
\end {cases}.
\end{equation}
Let us denote the ratio by $\mathcal{R}(N,T,k)=\lim_{N,T,k\rightarrow \infty}\frac{\mathcal{Z}^{(GWW)}_{CSM}}{\mathcal{Z}^{(GWW)}_{YM}}=\lim_{x\rightarrow \pm \infty} F(x)$, then by using Eq. (\ref{eq: leading asymp of F}) and putting $T=\zeta N, \ k=\lambda^{-1}N$, the ratio of the partition functions is given by,
\begin{equation}\label{eq: asymptotic ratio}
\mathcal{R}(N,T,k)=
\begin{cases}
1 -\mathcal{O}(e^{-N})&\ k> (1+\epsilon) \mu(N,T)\\
0 +\mathcal{O}(e^{-N^2})&\ k<(1-\epsilon)\mu(N,T)
\end{cases},
\end{equation}
where $\epsilon$ is a positive infinitesimal parameter. Equivalently, the above result, up to leading order, can be written separately for two phases,
\begin{equation}\label{eq: ratio 1,2}
\textit{for}\ N\geq T: \
\mathcal{R}(k,N,T)\approx
\begin{cases}
1&\  k> N+T\\
0& \ k< N+T
\end{cases},\hspace{.5cm}
\textit{for}\ N< T:\
\mathcal{R}(k,N,T)\approx
\begin{cases}
1& \ k> 2\sqrt{NT}\\
0& \  k< 2\sqrt{NT}
\end{cases}.
\end{equation}

Remembering that the $n-$th order of phase transition is defined by a discontinuity in the $n-$th derivative of free energy, thus, since the ratio of the partition functions jumps across the line $k=\mu$, one would expect a phase transition of order zero, however, more careful analysis by considering the sub-leading terms in the ratio is needed to determine the order of the phase transition via the explicit form of the free energy.  As it is observed, CSM theory has a complicated moduli space of three parameters ${N,T,k}$ indicating a phase transition at asymptotic regime.

In order to explain the above result for the ratio, we need to understand the phase structure of the eigenvalues of the holonomy matrix, see Appendix (A.1.2). 
By definition, density of the eigenvalues of the holonomy matrix is positive and normalized by $\int_{-\pi}^\pi\rho(\alpha)d\alpha=1$. The eigenvalue density is obviously restricted from below by zero lower bound. The lower gaps in the eigenvalue density are defined by $\{\alpha|\alpha\sim\alpha+2\pi, \rho(\alpha)=0\}$ and this leads to the lower-gap phase of the CSM theory. As it is explained in Appendix (A.1.2), and in \cite{Ta}, in the CSM theory the eigenvalues of the holonomy matrix are discrete and distance between two consecutive eigenvalues is $2\pi/k$. The discretization of the eigenvalues leads to an upper bound, $\frac{k}{2\pi }\times \frac{1}{N}=\frac{1}{2\pi \lambda}$, in the eigenvalue density, as
\begin{equation}
0\le \rho(\alpha) \le \frac{1}{2\pi \lambda}.
\end{equation}
The upper bound on the eigenvalue density leads to the upper-gap phase with upper gaps defined by $\{\alpha|\alpha\sim\alpha+2\pi, \rho(\alpha)=\frac{1}{2\pi\lambda}\}$.
The phase structure of the CSM theory defined by Eq. (\ref{eq: effective CSM partition function}), can be understood via the comparison of the competing forces in the theory. There are two competing factors in this problem: I) the attractive potential $V(U)$, of order $\mathcal{O}(N^2)$ at leading order with $T=\zeta N$, which tends to clump the eigenvalues $\alpha_i$'s and II) the repulsive force, of order $\mathcal{O}(N^2)$, from the measure $\mathcal{D}U$ or equivalently from the Vandermonde determinant which repels the eigenvalues, roughly because it vanishes when two eigenvalues coincide.
Roughly speaking, competition of the attractive force from the potential and a repulsive force from the Vandermonde determinant in the matrix integral, results to a phase transition. In fact, there are two phases; in one phase the attraction force is weaker than the repulsive force and the eigenvalues has support on the whole circle. In another phase, the repulsive force is weaker and this leads to the existence of a gap in the eigenvalue density and therefore the eigenvalues have only support on finite arcs on the circle.

In summary, the saddle point analysis in the variational problem (\ref{eq: variational problem}) indicates that in the CSM theory with the GWW potential, there are four phases, namely \textit{no-gap} phase, \textit{lower-gap} phase, \textit{upper-gap} phase and $\textit{two (lower and upper)-gap}$ phase. Depending on the parameters $\lambda$ and $\zeta$, the phases of the CSM theory are classified in \cite{Ta}. For summary of details of the phase structure of this model see the Appendix (A.1.2)

Finally, a possible interpretation of the above result (\ref{eq: ratio 1,2}) can be expressed in terms of the continuum limit of the CSM theory and/or existence of the upper-gap phase. Notice that the difference between the CSM theory and the YM theory is the presence of the discreteness and/or upper bound in eigenvalue density for the CSM theory. Therefore, the continuum limit of the CSM is expected in the regime or the phase that upper bound is not saturated. On the other hand, in the continuum limit the partition function of the CSM theory becomes the partition function of the YM theory and the ratio becomes one. Therefore, as we are interested in the phase structure of the theory at large $N,k,T$ with fixed $\zeta, \lambda$, from Eq. (\ref{eq: ratio 1,2}) with the change of variable in terms of $\lambda$ and $\zeta$ the ratio with the correct interpretation can be written as
\begin{equation}\label{eq: ratio 3,4}
\textit{for}\ \zeta \leq 1:\ \mathcal{R}(\lambda, \zeta)\approx
\begin{cases}
1& \ \textit{no upper-bound/continuum phase},\hspace{.5cm}  \lambda^{-1}> \zeta+1\\
0& \ \textit{upper-bound/discrete phase},\hspace{.5cm} \lambda^{-1}< \zeta+1
\end{cases},
\end{equation}
\begin{equation}
\textit{for}\ \zeta>1:\
\mathcal{R}(\lambda, \zeta)\approx
\begin{cases}
1& \ \textit{no upper-bound/continuum phase},\hspace{.5cm}  \lambda^{-1}> 2\sqrt{\zeta}\\
0& \ \textit{upper-bound/discrete phase},\hspace{.5cm} \lambda^{-1}< 2\sqrt{\zeta}
\end{cases}.
\end{equation}
In the next part, explicit calculations for the free energy in each region of the moduli space determine the order of this new phase transition and other features of the CSM theory.

Above interpretation is consistent with the summarized results in the Appendix (A.1.2). This consistency can be seen by comparing our above result with the rearranged form of (\ref{eq: CSM phases}),
\begin{equation}
\textit{for}\ \lambda<\frac{1}{2}:
\begin{cases}
\textit{no-gap or lower-gap},& \  \zeta<\frac{1}{4\lambda^2}\\
\textit{two-gap},& \ \zeta>\frac{1}{4\lambda^2}
\end {cases},\hspace{.2cm}
\textit{for}\ \lambda>\frac{1}{2}:
\begin{cases}
\textit{no-gap},&\ \zeta<\frac{1}{\lambda}-1\\
\textit{upper-gap or  two-gap},& \ \zeta>\frac{1}{\lambda}-1
\end {cases}.
\end{equation}
However, in contrast to \cite{Ta}, our phase structure result is completely based on the analysis of the ratio of partition functions, without calculating the eigenvalue density.  Moreover, as it will be clarified, our result contains more information about the phase structure such as the explicit formulas for the free energy which gives the orders of phase transitions between continuous and discrete YM theories (CSM theory) and also between discrete YM theories (CSM theories) in different phases.
The actual properties and features of these new phase transitions can only be described by careful analysis of the Tracy-Widom distribution which will be presented in the next part.
\subsection{Free energy of CSM theory and orders of phase transitions}
In this part, an explicit formula for the free energy of the CSM theory in different phases is obtained. The free energy determines the order of the new phase transition introduced in the previous part. The obtained results will be checked in nontrivial calculations. Furthermore, some limits of the free energy in interesting points of the moduli space are computed.

In order to study the phase structure of the theory, the first step is to compute the sub-leading corrections to ratio formula, Eq. (\ref{eq: asymptotic ratio}), by using the expansion of the Tracy-Widom distribution. In the asymptotic regime, the free energy of the YM and CSM theories is defined via their partition functions, $\mathcal{F}_{YM/CSM}=\lim_{N\rightarrow\infty}\frac {1}{N^2}\log Z_{YM/CSM}$. In the case of GWW potential, by using Eq. (\ref{eq: ratio for GW}), and the integral representation of the Tracy-Widom distribution (see Appendix B), the free energy of the CSM theory is given by
\begin{equation}\label{eq: CSM PF TW}
\mathcal{F}_{CSM}= \mathcal{F}_{YM}+\frac{1}{N^2}\log{F(x)}= \mathcal{F}_{YM}-\frac{1}{N^2}[\int_x^\infty(s-x)q^2(s)ds],
\end{equation}
where $x=\frac{k-\mu}{\sigma}$.
In different phases, by using YM free energy, \cite{Gr-Wi},
\begin{equation}\label{eq: Gross Witten free energy}
\mathcal{F}_{YM}^{(GWW)}(\zeta)=
\begin{cases}
\frac {\zeta^2}{4}& \  0<\zeta<1\\
\zeta-\frac {3}{4} -\frac {\log \zeta}{2}&\  \zeta>1
\end{cases},
\end{equation}
the free energy of the CSM theory is given by
\begin{equation}
\mathcal{F}_{CSM}=\mathcal{F}_{YM}+\frac{1}{N^2}\log{F(N^{\frac{2}{3}}j)}=
\begin{cases}
 \frac{\zeta^2}{4}-\frac{1}{N^2}[\int_{N^{\frac{2}{3}}j}^\infty(s-N^{\frac{2}{3}}j)q^2(s)ds]&\ \zeta\leq 1\\
 \zeta-\frac{3}{4}-\frac{\log\zeta}{2}-\frac{1}{N^2}[\int_{N^{\frac{2}{3}}j}^\infty(s-N^{\frac{2}{3}}j)q^2(s)ds]& \ \zeta>1
\end{cases},
\end{equation}
where
\begin{equation}
j:=
\begin{cases}
\frac {k-(N+T)}{ 2^{-\frac{1}{3}} T^{\frac {1}{3}}}N^{-\frac{2}{3}}=\frac{\lambda^{-1}-(\zeta+1)}{2^{-\frac{1}{3}}\zeta^{\frac{1}{3}}}&\ \zeta\leq 1\\
\frac {k-( 2\sqrt {NT} )}{ 2^{-\frac{2}{3}}T^{\frac {1}{3}}(\sqrt {\frac {N}{T}}+\sqrt {\frac {T}{N}})^{\frac {1}{3}}}N^{-\frac{2}{3}}=\frac{\lambda^{-1}-2\zeta^{\frac{1}{2}}} { 2^{-\frac{2}{3}}\zeta^{\frac{1}{3}}(\zeta^{\frac{1}{2}}+ \zeta^{-\frac{1}{2} })^{\frac{1}{3}}}& \ \zeta>1
\end{cases}.
\end{equation}
Following the interpretation of the ratio result in the previous section, Eq. (\ref{eq: CSM PF TW}) can be written as
\begin{equation}\label{eq: CSM free energy}
\mathcal{F}_{CSM}=\mathcal{F}_{YM}+\mathcal{F}_{ub},
\end{equation}
where $\mathcal{F}_{ub}=\frac{1}{N^2}\log{F(N^{\frac{2}{3}}j)}=-\frac{1}{N^2}\int_{N^{\frac{2}{3}}j}^{\infty}(s-N^{\frac{2}{3}}j )q^2(s)ds$ is the upper bound energy.
For $j<0$, in the large $N$ limit, by using the expansion of $q(s)$ in the limit $s\rightarrow\pm\infty$, (see Appendix B), the following approximate result up to leading order is obtained,
\begin{eqnarray}\label{eq: monopole free energy}
\mathcal{F}_{ub}&=&-\frac{1}{N^2}\int_{N^{\frac{2}{3}}j}^{\infty}(s-N^{\frac{2}{3}}j )q^2(s)ds\nonumber\\
&\approx&-\frac{1}{N^2}\int(s-N^{\frac{2}{3}}j )(\frac{-s}{2})ds\bigg\vert_{s=N^{\frac{2}{3}}j\rightarrow-\infty}\nonumber\\
&&-\frac{1}{N^2}\int(s-N^{\frac{2}{3}}j )q^2(s)ds\bigg\vert_{s=\textit{finite}}\nonumber\\
&&-\frac{1}{N^2}\int(s-N^{\frac{2}{3}}j )(-\frac { e^{-\frac {2}{3}s^{\frac {3}{2}}} }{2\sqrt {\pi}s^{\frac {1}{4}}})^2ds\bigg\vert_{s\rightarrow\infty}\nonumber\\
&\approx& \frac{j^3}{12},
\end{eqnarray}
where we separated the integrand in the first line into three terms for $s\rightarrow -\infty$, finite $s$ and $s\rightarrow +\infty$, since the integration in the limit $N\rightarrow +\infty$ is from minus infinity to plus infinity. Notice that $s$ is bounded from below by $N^{\frac{2}{3}}j$ and therefore it cannot tend to minus infinity faster than $N^{\frac{2}{3}}j$. The third and fourth lines vanish in the large $N$ limit and only the second line gives a finite contribution and the result can be obtained by evaluation of the integral in the second line. Above approximate calculation produce the correct result as it will be shown in the following.

In the next-to-leading order, the upper bound energy and total free energy of CSM theory in each phase can be computed, either by using the above result, Eq. (\ref{eq: monopole free energy}), or by direct computation from Eq. (\ref{eq: CSM free energy}), and the free energy of GWW model, Eq. (\ref{eq: Gross Witten free energy}), and a new obtained result for the expansion of the Tracy-Widom distribution, Eq. (\ref{eq: asymp of Tracy Widom}), \cite{Ba-Bu-Di}, and the result is
\begin{equation}
\textit{for}\ \zeta\leq 1:
\mathcal{F}_{CSM}=
\begin{cases}
\frac{\zeta^2}{4} +\frac{1}{N^2}\log(1-\frac{e^{-c_1 j^{3/2}N}}{32\pi j^{3/2}N})& \ \lambda^{-1}>\zeta+1\\
\frac{\zeta^2}{4}+\frac{1}{N^2}\log(c_3\frac{e^{-c_2 |j|^3 N^2}}{|j|^{1/8}N^{1/12}})&\ \lambda^{-1}<\zeta+1
\end{cases},
\end{equation}
where $c_1=4/3,c_2=1/12,c_3=2^{\frac{1}{42}}e^{(\zeta^*)'(-1)}$ and $j=\frac{\lambda^{-1}-(\zeta+1)}{2^{-\frac{1}{3}}\zeta^{\frac{1}{3}}}$ in this phase, and in the other phase,
\begin{equation}
\textit{for}\ \zeta> 1:
\mathcal{F}_{CSM}=
\begin{cases}
\zeta-\frac{3}{4}-\frac{\log\zeta}{2} +\frac{1}{N^2}\log(1-\frac{e^{-c_1 j^{3/2}N}}{32\pi j^{3/2}N})& \ \lambda^{-1}>2\zeta^{\frac{1}{2}}\\
\zeta-\frac{3}{4}-\frac{\log\zeta}{2} +\frac{1}{N^2}\log(c_3\frac{e^{-c_2 |j|^3 N^2}}{|j|^{1/8}N^{1/12}})&\ \lambda^{-1}< 2\zeta^{\frac{1}{2}}
\end{cases},
\end{equation}
where $j=\frac{\lambda^{-1}-2\zeta^{\frac{1}{2}}} { 2^{-\frac{2}{3}}\zeta^{\frac{1}{3}}(\zeta^{\frac{1}{2}}+ \zeta^{-\frac{1}{2} })^{\frac{1}{3}}}$.

In the large $N$ limit, by expanding the logarithm, the CSM free energy in leading order can be computed from above equations,
\begin{equation}\label{eq: CSM partition function one}
\textit{for}\ \zeta\leq 1:
\mathcal{F}_{CSM}=
\begin{cases}
\frac{\zeta^2}{4}&\  \lambda^{-1}>\zeta+1\\
\frac{\zeta^2}{4}-\frac{c_2}{2^{-1}\zeta}|{\lambda^{-1}-(\zeta+1)}|^3&\  \lambda^{-1}<\zeta+1
\end{cases},
\end{equation}

and
\begin{equation}\label{eq: CSM partiton function two}
\textit{for}\ \zeta>1:
\mathcal{F}_{CSM}=
\begin{cases}
\zeta-\frac{3}{4}-\frac{\log\zeta}{2}& \ \lambda^{-1}>2\zeta^{\frac{1}{2}}\\
\zeta-\frac{3}{4}-\frac{\log\zeta}{2} -\frac{c_2}{2^{-2} \zeta(\zeta^{\frac{1}{2}}+ \zeta^{-\frac{1}{2} })}|\lambda^{-1}-2\zeta^{\frac{1}{2}}|^3&\ \lambda^{-1}< 2\zeta^{\frac{1}{2}}
\end{cases}.
\end{equation}

Thus, in different regions of the moduli space, the upper bound energy is given by
\begin{equation}\label{eq: upper bound free energy}
\mathcal {F}_{ub} =
\begin{cases}
-\frac{c_2}{2^{-1}\zeta}|{\lambda^{-1}-(\zeta+1)}|^3&\ \zeta\leq 1, \ \lambda^{-1}<\zeta+1\\
-\frac{c_2}{2^{-2} \zeta(\zeta^{\frac{1}{2}}+ \zeta^{-\frac{1}{2} })}|\lambda^{-1}-2\zeta^{\frac{1}{2}}|^3& \  \zeta> 1, \lambda^{-1}< 2\zeta^{\frac{1}{2}}
\end{cases}.
\end{equation}
After inserting $c_2=1/12$, one can see that the above result is exactly the same as Eq. (\ref{eq: monopole free energy}). It would be interesting to compare the above explicit expressions for the free energy, and possible analytic expressions for free energy that would be obtained from Eq. (\ref{eq: variational free energy}) by putting the equilibrium eigenvalue density of CSM theory with GWW potential. However, by using Eq. (\ref{eq: variational free energy}), the general formula of free energy is used in the discussion of level-rank duality, in \cite{Ta} and \cite{Tak}, but in comparison to our studies, the analytic formula for free energy as a function of $\lambda$ and $\zeta$, is not obtained and consequently the orders of the phase transitions in the CSM theory with any potential such as GWW potential are not discussed in \cite{Ta} and \cite{Tak}. After some remarks, based on the obtained results we will discuss the phase structure and the orders of phase transitions in CSM theory.

It can be seen from Eq. (\ref{eq: upper bound free energy}) that, depending on the phase, the upper bound forms at a critical coupling $\lambda_c$, which can be determined exactly in each phase,
\begin{equation}
\lambda_c^{-1}=
\begin{cases}
2(1-\epsilon)\zeta^{\frac{1}{2}}&\  \zeta>1\\
(1-\epsilon)(1+\zeta)&\  \zeta\leq 1
\end{cases},
\end{equation}
and upper bound critical energy, $\mathcal{F}_{ub}^{crit}$, infinitesimally close to the domain wall, at $\lambda_c$, is
\begin{equation}
\mathcal{F}_{ub}^{crit}=
\begin{cases}
32c_2\epsilon^3\zeta^{\frac{3}{2}}/(\zeta^{\frac{3}{2}}+\zeta^{\frac{1}{2}})&\ \zeta>1\\
2c_2\epsilon^3(1+\zeta)^3/\zeta&\ \zeta\leq 1
\end{cases}.
\end{equation}

The free energy of CSM theory, Eqs. (\ref{eq: CSM partition function one}) and (\ref{eq: CSM partiton function two}), determines the phase structure of the CSM theory, Figure (1). In fact, the explicit formula for the free energy of the upper bound, determines the order of the phase transition between the continuous theory (YM theory) and discrete theory (CSM theory).
From the upper bound energy, Eq. (\ref{eq: upper bound free energy}),
it is easy to find that the phase transition at $\lambda^{-1}=\zeta+1$, between regions (I) and (II), and the phase transition at $\lambda^{-1}=2\zeta^{\frac{1}{2}}$, between regions (III) and (IV) are third order, i.e. the third derivative of the free energy, with respect to $\lambda$ or $\zeta$, jumps at the critical line.
By direct computation one can observe that a second-order phase transition between regions (I) and (IV) happens at domain wall $\zeta=1$. One needs to check the continuity of derivative of free energy, with respect to $\zeta$, in regions (I) and (IV) at $\zeta=1$ up to second derivative. However, this domain wall separates also regions (II) and (III) and in those regions, this is a third-order domain wall by simple computation from YM free energy, Eq. (\ref{eq: Gross Witten free energy}). Thus, upper bound changes the third-order domain wall to a second-order domain wall.
\begin{center}
\begin{figure}
\includegraphics[width=10cm]{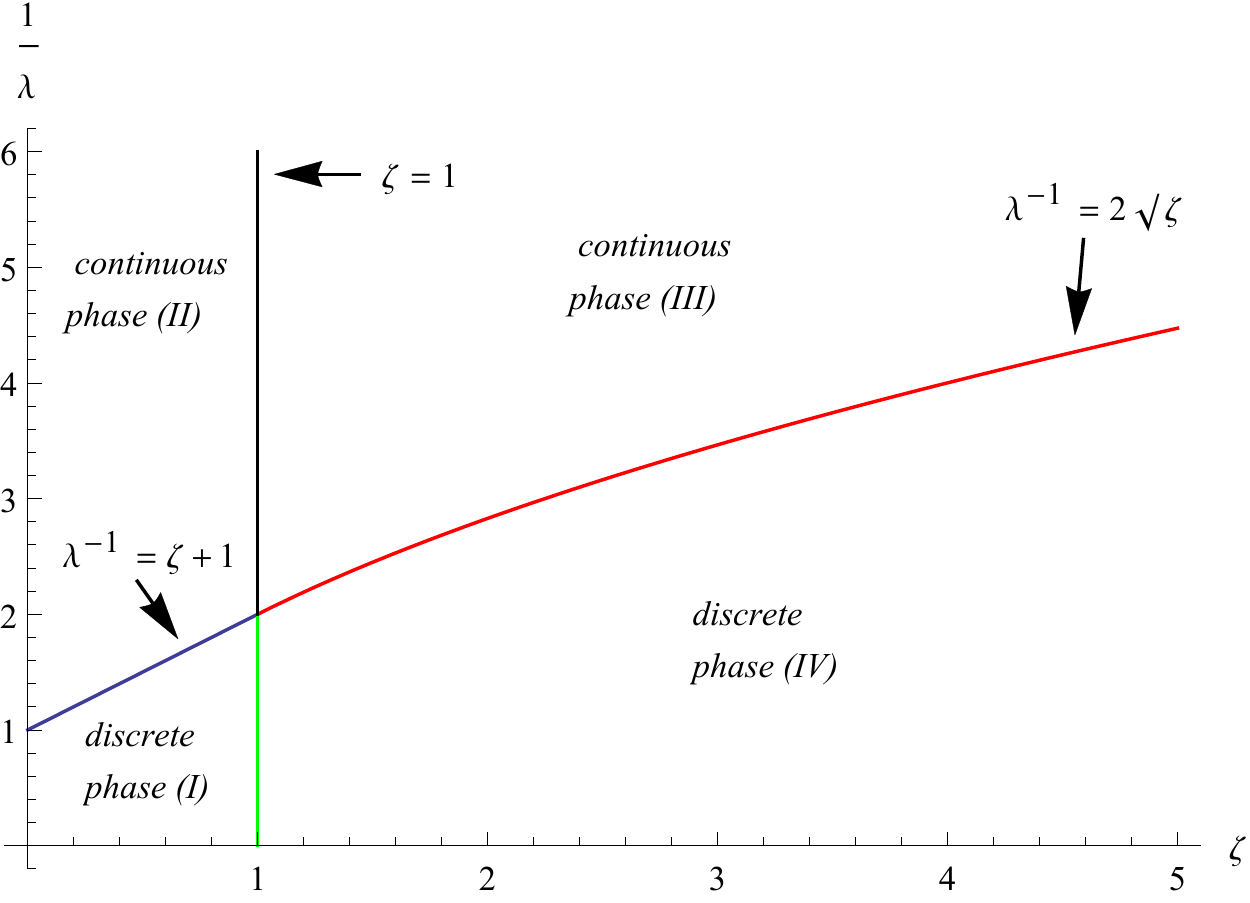}
\caption{Phase structure of Chern-Simons matter theory with GWW potential. Black line (separating regions II and III) and blue line (separating regions I and II), and red curve (separating regions III and IV) are third-order domain walls and green line (separating regions I and IV) is a second-order domain wall.}
\end{figure}
\end{center}
In summary, the above results in CSM theory with GWW potential determine the phase structure of the theory and as a new result of our study we have determined the order of phase transitions. In this theory, there are four different domain walls separating four different phases. I) At the GWW domain wall, the black line, $\zeta=1$, where a third-order phase transition occurs and II) at the green line, $\zeta=1$, separating two discrete theories, a second-order phase transition occurs, and III) two third-order domain walls separating continuous theories and discrete theories; the blue line, $\lambda^{-1}=\zeta+1$ in $\zeta\leq 1$ phase and the red curve, $\lambda^{-1}=2\zeta^{\frac{1}{2}}$ in $\zeta>1$ phase.
\subsection{Consistency checks and remarks}
Consistency checks can be observed in the following. The YM eigenvalue density, $\rho(\alpha)$, Eq. (\ref{eq: continuous measure}), is maximum at $\alpha=0$. Hence the upper bound condition, $\rho(\alpha) < \frac{1}{2\pi \lambda}$, becomes
\begin{equation}
\rho(\alpha=0)=
\begin{cases}
\frac1{2\pi} (1+\zeta \cos\alpha)\bigg\vert_{\alpha=0}=\frac{1}{2\pi}(1+\zeta) < \frac{1}{2\pi \lambda}& \  \zeta\leq 1\\
\frac{\zeta}{\pi} \cos(\alpha/2) \sqrt{\frac{1}{\zeta} -\sin^2(\alpha/2)}\bigg\vert_{\alpha=0}=\frac{\sqrt{\zeta}}{\pi} < \frac{1}{2\pi \lambda}&\ \zeta>1
\end{cases}.
\end{equation}
These conditions are precisely the conditions in $\mathcal{F}_{ub}=0$ or $\mathcal{R}(\lambda,\zeta)=1$, in regions (II) and (III). In this case, the CSM theory is described by the continuum YM theory and thus the eigenvalue density of the CSM theory, which in the phase $\mathcal{R}(\lambda,\zeta)=1$ is the eigenvalue density of YM theory, should satisfy the upper bound.

In the following, some explicit values of the upper bound energy in strong and weak coupling regime are computed.
Taking the strong coupling limit, $\lambda\rightarrow\infty$, is only possible in regions (I) and (IV) and the values of the upper bound energy at strong coupling are given by
\begin{equation}
\textit{in region (I)}:
\mathcal{F}_{ub}=
\begin{cases}
16c_2&\  \zeta\rightarrow 1\\
\infty&\  \zeta\rightarrow 0
\end{cases},
\end{equation}
and
\begin{equation}
\textit{in region (IV)}:
\mathcal{F}_{ub}=
\begin{cases}
16c_2&\  \zeta\rightarrow 1\\
32c_2&\  \zeta\rightarrow \infty
\end{cases}.
\end{equation}
More generally, in region (IV), for any fixed $\lambda$, $\mathcal{F}_{ub}=32c_2$ as $\zeta\rightarrow \infty$.
In the zero coupling limit (free limit), $\lambda\rightarrow 0$, we expect to see the continuous YM theory and thus $\mathcal{F}_{ub}=0$. This limit is only possible and meaningful in region (IV) (because in regions (II) and (III) the CSM theory is always in continuous phase) and in this region it implies $\zeta\rightarrow\infty$ which leads to $\mathcal{F}_{ub}\rightarrow 0$, as $\lambda\rightarrow 0$ and $\zeta\rightarrow\infty$, because $\lambda^{-1}-2\zeta^{\frac{1}{2}}\rightarrow 0$, as $\epsilon\rightarrow 0$. Such behavior in zero coupling limit is in fact consistent with results of other studies in \cite{Sh-Yi}, \cite{Gi} and \cite{Ch-Mi}. It has been shown in those studies that the thermal partition function of CSM theory in the free limit is governed by continuous YM partition function (\ref{eq: Matrix integral for Yang Mills}) and the phase structure of the theory is similar to that of YM theory.

On the line $\lambda=\frac{1}{2}$, in region (II), $\mathcal{F}_{ub}=0$ and in region (IV), $\mathcal{F}_{ub}=\frac{-32c_2}{\zeta(\zeta^{\frac{1}{2}}+ \zeta^{-\frac{1}{2} })}(1-\zeta^{\frac{1}{2}})^3$. At the point $\zeta=1, \lambda=\frac{1}{2}$; $\mathcal{F}_{ub}=0$ and at $\zeta\rightarrow\infty, \lambda=\frac{1}{2}$; $\mathcal{F}_{ub}=32c_2$.

\section{Interpretations and discussions}
There are some possible explanations as well as interpretations of our results. A possible explanation is via the understanding of the physical picture of the phase transition. The level $k$ is a formal source of discreteness in the system and the real source of discreteness is the existence of the upper bound. In fact, since all the parameters are going to infinity, taking the continuum limit strictly depends on the relations between the parameters. The plausible continuum limit exists under certain conditions. In the naive continuum limit, we expect to remove the upper bound, $\lambda\rightarrow 0$, on the eigenvalue density, but at the same time, for larger and larger $T$ the eigenvalue density is sharper and peak of the eigenvalue density is higher, so, $\lambda\rightarrow 0$ limit should be taken before the limit $T\rightarrow \infty$. On the other hand, for finite $\lambda$ in some phases, although the eigenvalue density is bounded but the upper bound is not saturated and the upper-gap is not formed and therefore the continuum limit of the CSM theory exists. Therefore, we can explain our results as a sufficient and necessary conditions for the continuum limit. In summary, in order to get the expected continuum limit, $k$ should go to infinity faster than $N$ (and $T$), i.e. the relations $k>N+T$ and $k>2\sqrt{NT}$ in two phases.

In this study we have observed a new second order phase transition between two CSM theories in regions (I) and (IV). It would be interesting to understand possible physical meanings and implications of this second order domain wall.

In the upper-gap phases, the partition function and free energy of the CSM theory, in addition to the YM part, contain an extra term originated from the upper bound. The upper gap is a consequence of discretization of the eigenvalues of the holonomy matrix and the discreteness is itself a direct consequence of the summation over flux sectors in the partition function. Discrete fluxes are dual to the monopole operators and discrete flux configurations are effectively the same as the configurations generated by monopoles at the centre of $S^2$. However, in our case, there is no physical monopole or any matter inside $S^2$. Having said that, we might interpret the formation of the upper bound configuration and its energy $\mathcal{F}_{ub}$ as the formation of the monopole in the CSM theory with the same energy. Possible physical implications of our results in the context of monopoles, e.g. monopole condensation and confinement is highly interesting and remain for future studies.

As it is discussed in \cite{Ta} and \cite{Tak} and references therein, the level-rank duality in CSM theory leads to conservation of the partition function under the transformation $N\rightarrow k-N$ and $k\rightarrow k$ or equivalently, $\lambda\rightarrow 1-\lambda$ and $\zeta\rightarrow \zeta(\lambda/(1-\lambda))$.
At $\lambda=\lambda_c=1/2$, the level-rank duality is valid. But in the general case, the level-rank duality does not seem to be valid. At $\lambda=\lambda_c$, the CSM free energy in each phase satisfies the level-rank duality but further studies are needed to find other solutions, $\lambda^*$, satisfying,
\begin{equation}\label{eq: level rank duality}
\mathcal{F}_{CSM}(\lambda, \zeta)\bigg\vert_{\lambda^*}=\mathcal{F}_{CSM}(1-\lambda, \frac{\zeta\lambda}{1-\lambda})\bigg\vert_{\lambda^*}.
\end{equation}

A similar third-order phase transition, between discrete and continuous two-dimensional gauge theories, occurs in the Douglas-Kazakov model, \cite{Do-Ka}. Similar phase transitions happen also in $2d$ $q-$deformed YM theory, \cite{Ar}, \cite{Ca} and \cite{Ja-Ma}. It would be interesting to explore possible relations between our study and these works.

In this paper, the CSM theory is considered with the GWW potential, however, the general results in section (3.1) allow to study the asymptotic limit of the ratio of the partition functions of CSM and YM theories for any arbitrary potential. One would expect to get similar phase structure of upper-gap and no-upper-gap phases for an arbitrary nontrivial potential. We conjecture that the orders of phase transitions between different phases of CSM theory with any potential in Fig. (1) is universal second and third orders. However, the actual computations of the free energy and the order of phase transitions remain for future studies.

Possible generalizations of the obtained results in this paper, e.g. the third-order phase transitions, to CSM theory with other gauge groups $G=O(N)$ and $Sp(N)$, remain for future studies, \cite{Ch-Jo}.

Another direction for further study is possible applications and implications of our results, especially the new phase structure of the model, in AdS/CFT correspondence with Vasiliev higher spin gravity.

\vspace{.5cm}
\thanks{\textbf{Acknowledgements.}} Thanks to N. Jokela and T. Takimi for useful discussions and reading the manuscript.

\begin{appendices}
\section{Random matrix theory, Toeplitz determinant and Riemann-Hilbert problem in gauge theory}\label{App: AppendixA}
In this appendix, two-dimensional Yang-Mills theory and three-dimensional Chern-Simons matter theory, and their relation to matrix models and Toeplitz determinants are reviewed. The examples of YM and CSM theories with the Gross-Witten-Wadia potential are considered and explicit formulas for the partition function and free energy are reviewed. Furthermore, the phase structure of the models and the eigenvalue distribution are described. Finally, the GWW model as the solution of the Riemann-Hilbert problem is briefly stated.
\subsection{YM theory, CSM theory, matrix models and Toeplitz determinant}
\subsubsection*{YM theory and CSM theory as matrix models}
The interesting and rich phase structure of $2d$ YM theory and its matrix model representation with some special potentials have been studied in both mathematics and physics literature, \cite{Gr-Wi}, \cite{Wa} and \cite{Ba-De-Jo}.
The two-dimensional YM theory is defined via the following matrix integral representation of the partition function,
\begin{eqnarray}\label{eq: Matrix integral for Yang Mills}
\mathcal{Z}_{YM}=\int [\mathcal{D}U]e^{-V_{YM}}&=&\prod_{j=1}^N\int_{-\infty}^\infty d\alpha_j\prod_{l< p} \left(2\sin (\frac {\alpha_l-\alpha_p}{2})\right)^2 e^{-V_{YM}}\nonumber\\
&=&\prod_{j=1}^N\int_{-\infty}^\infty d\alpha_j\prod_{l< p} |e^{\ii\alpha_l}-e^{\ii\alpha_p}|^2 e^{-V_{YM}},
\end{eqnarray}
where $V_{YM}$ is the potential of the YM theory. This is a partition function of coulomb gas of unit charges on the unit circle with an attractive potential $V_{YM}$ and a logarithmic repulsive potential, $\log|e^{\ii\alpha_l}-e^{\ii\alpha_p}|$, from the Vandermonde determinant.

Similar to the above result, the path integral of the CSM partition function Eq. (\ref{eq: CSM effective partition function}) can be evaluated by using the techniques from \cite{Bl-Th} for pure CS and its modification to include the effect of potential, \cite{Ta}. For the study become self contained, we summarize the steps of such computation and sketch the derivation of each step but for further details and implications we refer to the mentioned references. As we observed in section (2), after integrating out the massive fields in the path integral (\ref{eq: CSM partition function}), we are left with a path integral (\ref{eq: CSM effective partition function}) over massless fields and the holonomy field. In the following, we integrate on the remaining massless fields and we reduce the path integral to a matrix integral over the holonomy fields.

The first step towards this goal is fixing the gauge invariance of the theory. Let us write $A$ as $A_{1,2}$ along the $S^2$ direction and $A_3$ along the $S^1$ direction. Then, there are three gauge fixing: i) temporal gauge, $\partial_3 U=0$ for $U=e^{\beta A_3}$ which is the closest gauge to $A_3=0$ and leaves a $2d$ gauge invariance. For further constraining this gauge freedom, we impose ii) diagonalize $U(x)$ simultaneously for all $x\in S^2$. This constraint Abelianize the $2d$ gauge invariance to a $2d$ $U(1)^N$ gauge invariance. We further constrain the residual $2d$ Abelian gauge invariance by imposing iii) a Coulomb gauge on the time independent diagonal elements of $A_{1,2}$.

As we mentioned, in order to evaluate the path integral in (\ref{eq: CSM effective partition function}), we want to integrate over the remaining fields $A_{1,2}$ and reduce the path integral to a matrix integral over $U$. To do so, we impose the above gauge fixing and after some straightforward computations the path integral (\ref{eq: CSM effective partition function}) reduces to
\begin{equation}
\mathcal{Z}_{CSM}=\int[\mathcal{D}A_3][\mathcal{D}A_2][\mathcal{D}A_1]\Delta_{FP}e^{-\ii\frac {k}{2\pi}\int d^3x Tr (A_3(\partial_1A_2-\partial_2A_1)+D_3A_1A_2)-T\int d^2x\sqrt {g}v (U (x))},
\end{equation}
where $\Delta_{FP}$ is the Fadeev-Popov determinant of gauge fixing conditions and $D_3=\partial_3+[A_3,]$.

Now by using the Gaussian integration and Lagrange multiplier method it is straightforward to integrate over $A_{1,2}$ in the following manner. Notice that, because of the gauge fixing, the first term in the exponential receives contributions only from time independent diagonal elements of $A_{1,2}$ and these elements do not contribute to the second term in the exponential. Then, obviously the path integral over off-diagonal elements of $A_{1,2}$ in the second term is of the Gaussian form and yields $\frac{1}{\sqrt{Det_V (D_3)}}$, where $Det_V(D_3)$ is a determinant of the operator $D_3$ acting in wedge product over vector functions.

By imposing the gauge condition (iii), the path integrals over time-independent diagonal elements of $A_{1,2}$ in the first term of the exponential, evaluated to a delta function that fixes $U(x)$ to be constant on $S^2$. Then, because of the delta function, the path integral over the eigenvalues of the holonomy matrix is reduced to an ordinary integral over the constant eigenvalues of holonomy matrix. Lets $m_j$'s be constant units of flux for the $j-$th $U(1)$ factor, then after summing over all flux sectors, the full path integral for holonomy matrix reduces to,
\begin{equation}
  \prod_{j=1}^N\int_{-\infty}^\infty d\alpha_j\sum_{m_j=-\infty}^\infty e^{\ii k m_j\alpha_j}.
\end{equation}
As we will see, the above contribution from the $U(1)^N$ flux sectors leads to the discretization of the holonomy eigenvalues.

It is heuristically argued in \cite{Ta} and carefully evaluated in \cite{Bl-Th} that the Fadeev-Popov determinant of the gauge conditions is $\Delta_{FP}=Det_S(D_3)$, a determinant of the operator $D_3$ acting on scalar functions on $S^2$ and also $\sqrt{Det_V(D_3)}=Det'_S(D_3)$, which is a determinant of the operator $D_3$ acting on scalar functions on $S^2$ excluding the zero mode of the scalar field. Therefore, the Fadeev-Popov determinant mostly cancels by the path integral of the off-diagonal elements and the ratio of Fadeev-Popov determinant and determinant from the path integral over off-diagonal modes, which is independent of the flux sector, becomes
\begin{equation}
\frac{\Delta_{FP}}{\sqrt{\det_V(D_3)}}=\frac{\det_S(D_3)}{\det'_S(D_3)}=\prod_{l< p} \left(2\sin (\frac {\alpha_l (n_l) -\alpha_p (n_p) }{2})\right)^2.
\end{equation}
Putting above pieces of calculation together, partition function of CSM theory, Eq. (\ref {eq: CSM effective partition function}), is given by a discrete unitary matrix integral,
\begin{eqnarray}
\mathcal{Z}_{CSM}&=& \int [\mathcal{D}A] \ e^{(\ii \mathcal{S}_{CS}- T\int d^2x\sqrt {g}v (U (x)))}\nonumber\\
&=& \prod_{j=1}^N\int_{-\infty}^\infty d\alpha_j[\prod_{l< p} \left(2\sin (\frac {\alpha_l (\overrightarrow {n}) -\alpha_p (\overrightarrow {n}) }{2})\right)^2e^{-N\zeta v(U)}]\sum_{m_j=-\infty}^\infty e^{\ii k m_j\alpha_j}\nonumber\\
&=& \prod_{j=1}^N\int_{-\infty}^\infty d\alpha_j[\prod_{l< p} \left(2\sin(\frac {\alpha_l (\overrightarrow {n}) -\alpha_p (\overrightarrow {n}) }{2})\right)^2e^{-N\zeta v(U)}]\sum_{n\in\mathds{Z}}\delta(k\alpha_j-2\pi n)\nonumber\\
&=&\prod_{i=1}^N\sum_{n_i=-\infty}^\infty [\prod_{l< i} \left(2\sin(\frac {\alpha_l(\overrightarrow {n})-\alpha_i(\overrightarrow{n})}{2})\right)^2e^{-N\zeta v(U)}],
\end{eqnarray}
where from the second line to the third line we used $\sum_{m_j=-\infty}^\infty e^{\ii k m_j\alpha_j}=\sum_{n\in\mathds{Z}}\delta(k\alpha_j-2\pi n)$, and the delta function puts restriction on eigenvalues, $ \alpha_j( \overrightarrow{n})=\frac{2\pi n_j}{k}$, such that no two $n_i$ are allowed to be equal, which in return transform the integral over the eigenvalues to a discrete sum in the last line.

\subsubsection{Toeplitz determinant, matrix model and Brownian motion}
The relation between matrix integral and Toeplitz determinant is briefly introduced in the introduction.
In this part, in order to elaborate more on Eq. (\ref{eq: PaFu and ToDe}), the continuous and  discrete Toeplitz determinants are defined and their relations to continuous and discrete matrix models are stated.

A continuous Toeplitz determinant with a weight function $f$ is defined by
\begin {equation}
\mathrm{D}_N (f) =\det\left[\int_{|z|=1}z^{-j+l}f(z)\frac{dz}{2\pi\ii z}\right]_{j, l=0}^{N-1},
\end {equation}
where $z=e^{\ii\alpha}$.
Similarly, the discrete Toeplitz determinant is defined in a finite domain $d$ with the size $|d|$ as follows
\begin{equation}
\mathrm{D}_N^{|d|} (f, d)=\det\left[\frac {1}{|d|}\sum_{z\in d}z^{-j+l} f (z)\right]_{j, l=0}^{N-1}.
\end {equation}

The discrete and continuous Toeplitz determinants are related to the discrete and continuous matrix integrals via a relation called, Heine-Szeg\"{o} identity,
\begin{eqnarray}\label{eq: Heine-Szego}
\mathrm{D}_N (f) &=& \prod_{j=1}^N\int_0^{2\pi}\frac {d\alpha_j}{2\pi} f (e^{\ii\alpha_j})\prod_{l<p}| e^{\ii\alpha_l} - e^{\ii\alpha_p} |^2,\nonumber\\
\mathrm{D}_N^{|d|} (f, d)&=& \frac {1}{N! |d|^N}\sum_{(z_1, ..., z_N)\in d^N}\prod_{j=1}^N f (z_j) \prod_{1<j< l \leq N}|z_j-z_l|^2.
\end{eqnarray}

Therefore, the partition functions of YM theory and CSM theory are equal to the determinant of $N\times N$ continuous Toeplitz matrix (continuous Toeplitz determinant), $\mathrm{D}_N(f)$, and discrete Toeplitz determinants, $\mathrm{D}_N^{|d_1|} (f, d_1)$, on domain $d_1=\{z\in \mathds{C}| z^k=1\}$ with size $|d_1|=k$, respectively,
\begin{equation}\label{eq: PaFu and ToDe}
\mathcal{Z}_{YM}=\mathrm{D}_N(f),\hspace{.5cm} \mathcal{Z}_{CSM}=\mathrm{D}_N^k(f, d_1),
\end{equation}
where $f=e^{-V_{YM}}$ is the generating (weight) function.

In \cite{Ba-Li} discrete Toeplitz determinants are used to study non-intersecting Brownian motion. The main result of that study is the following result, Theorem (1.1) in \cite{Ba-Li}. Let $d$ be a finite discrete subset of a unit circle $S^1$ with size $|d|$ and let $\Omega$ be a neighborhood of $S^1$ and $f(z)$ be an analytic positive weight function with continuous measure, $f(z)\frac{dz}{2\pi \ii z}$, then,
\begin{equation}\label{ratio Fredholm}
\frac{\mathrm{D}_N^{|d|} (f, d)}{\mathrm{D}_N (f)}=\det {\left(1+K\right)},
\end{equation}
where $K$ is the integral operator with kernel $K(z,w)$ of the Fredholm determinant. And as we briefly pointed out, the kernel is given by a factor of continuous kernel as
\begin{equation}\label{eq: kernel}
K (z, w)=K_{cont}(z, w)\sqrt {v (z)v (w)f (z)f (w)},\hspace {.5cm} K_{cont}(z, w)=z^{-N}\frac {p_N (z)p_N^*(w)- p_N^* (z)p_N(w)}{1-z^{-1}w},
\end{equation}
and $p_N(z)$ is an orthogonal polynomial with respect to weight function $f(z)$, and $p^*_N(z):=z^N \overline{p_N(\bar{z}^{-1})}$, and $v(z)$ is a function representing discreteness. This function is defined by
\begin{equation}\label{general discrete function}
v(z):=
\begin{cases}
-\frac{z\gamma'(z)}{k\gamma(z)}&  \ z\in S^1_{\textit{in}}\\
\frac{z\gamma'(z)}{k\gamma(z)}-1&  \ z\in S^1_{\textit{out}}
\end{cases},
\end{equation}
where $\gamma(z)$ is an analytic function on $\Omega$ such that it vanishes exactly on $d$ and all zeros are simple, and $S^1_{\textit{in}}$ and $S^1_{\textit{out}}$ are positively-oriented circles of radius $1-\epsilon$ and $1+\epsilon$, for small $\epsilon > 0$. For proof of the above result in a slightly different notation, see \cite{Ba-Li}.

Toeplitz determinant appears naturally in the context of Brownian motion as probability. In order to state some other useful and relevant results from \cite{Ba-Li} which we are going to use them in gauge theory, let us briefly introduce a minimal necessary notation in random walk theory. Let $X_i(t)$, $i=1, ...,  N$ be independent standard Brownian motion with conditions, $X_0(t)<X_1(t)<...<X_{N-1}(t)$ for all $t\in [0,T]$ and $X_i(0)=X_i(T)$ and define the width by $W_N(T)=\sup_{t\in[0,T]}(X_{N-1}(t)-X_1(t))$. Then, the conditional probability for the width $W_N(T)$ of non-intersecting continuous time symmetric simple random walks is obtained in (Proposition 4.1), \cite{Ba-Li},
\begin{equation}\label{eq: conditional probability}
  \mathds{P}(W_N(T)<M)= \lim_{N,T,k\rightarrow\infty} \oint_{|s|=1}\frac{\mathrm{D}_N^{|d_s|}(f_{GWW},d_s)}{\mathrm{D}_N(f_{GWW})}\frac{ds}{2\pi\ii s},
\end{equation}
where $d_s= \{z\in \mathbb{C}|\ z^k=s\}$, and the asymptotic limit is obtained in (Theorem 4.1), \cite{Ba-Li},
\begin{equation}\label{eq: asymptotics of conditional probability}
  \lim_{N,T\rightarrow \infty}\mathds{P}(\frac{W_N(T)-\mu(N,T)}{\sigma(N,T)}\leq x)= F(x),
\end{equation}
where
\begin{equation}
\mu :=
\begin{cases}
N+T&  \ N\geq T\\
2\sqrt {NT}& \ N<T
\end{cases},
\hspace{.5 cm}
\sigma :=
\begin{cases}
 2^{-\frac{1}{3}} T^{\frac {1}{3}}& \  N\geq T\\
 2^{-\frac{2}{3}}T^{\frac {1}{3}}(\sqrt {\frac {N}{T}}+\sqrt {\frac {T}{N}})^{\frac {1}{3}}& \  N<T
\end{cases},
\end{equation}
and $F(x)$ is the Tracy-Widom distribution, discussed in Appendix B.

\subsubsection{Free energy and eigenvalue distribution function}
In the YM theory, in order to find the equilibrium measure one has to look for the measure on the circle which is non-negative and satisfies an energy minimization problem.

The distribution of the eigenvalues on the unit circle is determined by solving the following variational problem, \cite{Ba-De-Jo}.
For a given measure $f(z)\frac{dz}{2\pi\ii z}$, the equilibrium measure (eigenvalue density) $\rho(z)$ is obtained as a unique minimizer of the free energy in following variational problem,
\begin{equation}\label{eq: variational problem}
E^V=\inf\{\mathcal{F}^V((\rho)): \rho\ \textit{is a probability measure on the unit circle}
\ S^1\},
\end{equation}
where
\begin{equation}\label{eq: variational free energy}
\mathcal{F}^V(\rho)=\int\int_{S^1\times S^1}\log|z-w|^{-1}d\rho(z)d\rho(w)+\int_{S^1} V(z)d\rho(z),
\end{equation}
is the free energy.
\subsubsection*{Gross-Witten-Wadia model}
Consider a specific potential $V_{YM}=-(T/2) Tr(U+U^\dag)$ where $U$ is a $N\times N$ holonomy matrix with eigenvalues $\alpha_i$'s. This defines an interesting two-dimensional YM theory, called Gross-Witten-Wadia model, \cite{Gr-Wi}, \cite{Wa}, which is a unitary random one-matrix model. The partition function reads,
\begin{eqnarray}
\mathcal{Z}_{YM}^{(GWW)}&=&\int [\mathcal{D}U]e^{\frac{T}{2}Tr(U+U^\dag)}\nonumber\\
&=&\prod_{j=1}^N\int_{-\infty}^\infty d\alpha_j\prod_{l< p} \left(2\sin (\frac {\alpha_l-\alpha_p}{2})\right)^2 \prod_{j=1}^N e^{\frac{T}{2}(z_j+z_j^{-1})}\nonumber\\
&=&\prod_{j=1}^N\int_{-\infty}^\infty d\alpha_j\prod_{l< p} |e^{\ii\alpha_l}-e^{\ii\alpha_p}|^2 e^{T\sum_{j=1}^{N}\cos\alpha_j}.
\end{eqnarray}
The above partition function is equivalently given by the Toeplitz determinant, $\mathrm{D}_N(f)$ with $f=e^{T\cos\alpha}$.

By solving the variational problem for the GWW model, $V(z)=\frac{T}{2}(z+z^{-1})$, the eigenvalue density is obtained,
\begin{equation}\label{eq: continuous measure}
\rho(\alpha)=
\begin {cases}
\frac1{2\pi} (1+\zeta \cos\alpha)& \  \zeta\leq 1\\
\frac{\zeta}{\pi} \cos(\alpha/2) \sqrt{\frac{1}{\zeta} -\sin^2(\alpha/2)}&  \ \zeta>1
\end {cases}.
\end{equation}
As it will be explained in below, this model possesses a third-order phase transition and the phase transition happens at $\zeta=1$, and the eigenvalue density $\rho(\alpha)$ has different features in different phases. In $\zeta\leq 1$ phase, $\rho(\alpha)$ is supported on whole circle and this phase is called \textit{no-gap} phase. In $\zeta>1$ phase, $\rho(\alpha)$ is supported on a subset of circle, $[-\alpha_c, \alpha_c]$ where $\alpha_c$ is given by the condition $\sin^2(\alpha_c/2)=1/\zeta$ and this phase is called \textit{lower-gap} phase.
Free energy of the GWW model is defined by $\mathcal{F}_{YM}^{(GWW)}(\zeta)=\lim_{N\rightarrow\infty}\frac{1}{N^2}\log\mathcal{Z}_{YM}^{(GWW)}(e^{\zeta N\sum_j\cos\alpha_j})$, and can be computed by putting the eigenvalue density (\ref{eq: continuous measure}) in (\ref{eq: variational free energy}) and with a change of variable $T=\zeta N$, it has been computed in \cite{Gr-Wi},
\begin{equation}
\mathcal{F}_{YM}^{(GWW)}(\zeta)=
\begin{cases}
\frac {\zeta^2}{4}& \  0<\zeta<1\\
\zeta-\frac {3}{4} -\frac {\log \zeta}{2}&\  \zeta>1
\end{cases}.
\end{equation}
A third-order phase transition, i.e. discontinuity in $d^3\mathcal{F}/d^3\zeta$, occurs at $\zeta=1$.

\subsubsection*{CSM theory with GWW potential}
In the following, the qualitative and quantitative results for the phase structure of  CSM theory with GWW potential are reviewed. The explicit results are obtained from the saddle point analysis of the variational problem for the discrete GWW model, in low-temperature \cite{Ta} and high-temperature \cite{Tak}. For alternative derivations of the eigenvalue measures for models of this type, see \cite{Jo-Ja-Ke}.

At low temperatures, in the regions ($\lambda<1/2$, $\zeta<1$) and ($\lambda>1/2$, $\zeta<1/\lambda-1$), the repulsive force (factor (II) in discussion in section (3.2)) is stronger and eigenvalue density has support everywhere on the unit circle. The eigenvalue density is given by $\rho(\alpha)=(1+\zeta\cos\alpha)/2\pi$ and the system is in no-gap phase. As the temperature increases, the attractive force (factor (I)) plays more effective role and depending on $\lambda$, the lower-gap or the upper-gap forms first. For small $\lambda$, $\lambda<1/2$, in the region $1<\zeta<1/4\lambda^2$, the upper bound is large and first the lower-gap phase forms with eigenvalue density, $\rho(\alpha)= \zeta\cos(\alpha/2)/\pi \sqrt{1/\zeta-\sin^2(\alpha/2)}$ supported on  $[-\alpha_c, \alpha_c]$ and $\alpha_c$ is given by the condition $\sin^2(\alpha_c/2)=1/\zeta$. Then, in the region $\zeta>1/4\lambda^2$ the two-gap phase with one lower-gap and one upper-gap forms. The lower-gap and the upper-gap are located on the arcs $[e^{\ii b}, e^{-\ii b}]$ and $[e^{\ii a}, e^{-\ii a}]$ on the unit circle, respectively, where $a$ and $b$ are determined by
\begin{eqnarray}
\frac{1}{4\pi\lambda}\int_{-a}^{a}d\alpha\frac{1}{\sqrt{\sin^2{\frac{a}{2}}-\sin^2{\frac{\alpha}{2}}}\sqrt{\sin^2{\frac{b}{2}}-\sin^2{\frac{\alpha}{2}}}}&=&\zeta,\nonumber\\
\frac{1}{4\pi\lambda}\int_{-a}^{a}d\alpha\frac{\cos\alpha}{\sqrt{\sin^2{\frac{a}{2}}-\sin^2{\frac{\alpha}{2}}}\sqrt{\sin^2{\frac{b}{2}}-\sin^2{\frac{\alpha}{2}}}}&=&1+\frac{\zeta}{2}(\cos a+\cos b).
\end{eqnarray}
Between the arcs, in the complement of two gaps, the eigenvalue density is determined by
\begin{eqnarray}
\rho(\alpha)&=&\frac{|\sin\alpha|}{4\pi^2\lambda}\sqrt{(\sin^2{\frac{\alpha}{2}}-\sin^2{\frac{a}{2}})(\sin^2{\frac{b}{2}}-\sin^2{\frac{\alpha}{2}})}\nonumber\\
&\times&\int_{-a}^{a}d\theta\frac{1}{(\cos\theta-\cos\alpha)\sqrt{\sin^2{\frac{a}{2}}-\sin^2{\frac{\theta}{2}}}\sqrt{\sin^2{\frac{b}{2}}-\sin^2{\frac{\theta}{2}}}}.
\end{eqnarray}
For large $\lambda$, $\lambda>1/2$, in the region $1/\lambda-1<\zeta<1/(4\lambda(1-\lambda))$, the upper bound is small and upper-gap phase forms first, with eigenvalue density,
\begin{equation}
\rho(\alpha)=
\begin{cases}
\frac{1}{2\pi\lambda}-\frac{\zeta}{\pi}|\sin\frac{\alpha}{2}|\sqrt{\frac{1-\lambda}{\lambda\zeta}-\cos^2{\frac{\alpha}{2}}}& \  \cos^2{\frac{\alpha}{2}}<\frac{1-\lambda}{\lambda\zeta}\\
\frac{1}{2\pi\lambda}& \ \cos^2{\frac{\alpha}{2}}>\frac{1-\lambda}{\lambda\zeta}
\end {cases},
\end{equation}
then, in the region $\zeta>1/(4\lambda(1-\lambda))$, the two-gap phase forms.

In summary, the phase structure of the CSM theory with GWW potential is
\begin{equation}\label{eq: CSM phases}
\text{for}\ \lambda<\lambda_c:
\begin{cases}
\textit{no-gap phase} & \  \zeta<1\\
\textit{lower-gap phase}& \  1<\zeta<\frac{1}{4\lambda^2}\\
\textit{two-gap phase}& \ \zeta>\frac{1}{4\lambda^2}
\end {cases},\hspace{.2cm}
\text{for}\ \lambda>\lambda_c:
\begin{cases}
\textit{no-gap phase}&\  \zeta<\frac{1}{\lambda}-1\\
\textit{upper-gap phase}&\  \frac{1}{\lambda}-1<\zeta<\frac{1}{4\lambda(1-\lambda)}\\
\textit{two-gap phase}& \  \zeta>\frac{1}{4\lambda(1-\lambda)}
\end {cases},
\end{equation}
where $\lambda_c=\frac{1}{2}$. In principle, the above results for eigenvalue density can be put in Eq. (\ref{eq: variational free energy}) to give the free energy of GWW model. The general form of free energy, Eq. (\ref{eq: variational free energy}), has been discussed and used for CSM theory with different potentials including the GWW potential in the discussion of level-rank duality, \cite{Tak}. However, our new explicit results for free energy of the CSM theory with GWW potential as an analytic function of $\lambda$ and $\zeta$ is obtained from a completely different method.
\subsection{Riemann-Hilbert problem in Yang-Mills theory}
In this part, the YM theory as a continuous matrix model is formulated via a well-defined mathematical problem, namely the Riemann-Hilbert problem (RHP), \cite{Ba-De-Jo}. In the following, orthogonal polynomials and RHP associated to the GWW model are reviewed.
Lets $z\in \mathds{C}$, and $T\in \mathds{R}$ is a parameter of the model,
then the following polynomial,
\begin{equation}
p_N(z,T)=a_N(T)z^N+...
\end{equation}
is called an orthogonal polynomial with respect to weight function, $f(e^{\ii\alpha})=\exp{((T/2)(e^{\ii\alpha}+e^{-\ii\alpha}))}$, on the unit circle if it satisfies
\begin{equation}\label{orthogonal polynomials}
\int_{-\pi}^{\pi} p_N(e^{\ii\alpha})\overline{p_M(e^{\ii\alpha})}d\mu_f=\delta_{N,M},
\end{equation}
where $N,M>0$ and measure is $d\mu_f=f(e^{\ii\alpha})\frac{d\alpha}{2\pi}$.
The well-known connection between the orthogonal polynomial and Toeplitz determinant is established in \cite{Sz}. As a result, the leading coefficient in orthogonal polynomial, $a_N(T)$, is expressed in terms of the Toeplitz determinant,
\begin{equation}\label{eq: OP and TD}
a_N(T)=\frac{\mathrm{D}_{N-1}(T)}{\mathrm{D}_{N}(T)},
\end{equation}
where $\mathrm{D}_{N}(T)=\mathrm{D}_{N}(e^{T\cos\alpha})$.

It has been observed earlier that the partition function of the YM theory with the GWW potential which is given by a unitary matrix integral is expressed in terms of the Toeplitz determinant. Thus using the above connection to the orthogonal polynomials and the Szeg\"{o} strong limit, $\lim_{N\rightarrow 1}\mathrm{D}_{N}(T)=1$, one can obtain,
\begin{equation}
\log\mathcal{Z}_{YM}^{(GGW)}(T)=\log\mathrm{D}_{N-1}(e^{T\cos \alpha})= \sum_{l=N}^{\infty}\log a_l^2(T).
\end{equation}
To control the behavior of the partition function in the limit of large $N,T$, one needs to control the behavior of orthogonal polynomial $a_l^2(T)$ in the large $N,T$ limit for all $l\geq N$. To this aim, the steepest descent methods has been used to compute the asymptotic behavior of the following RHP, \cite{De-Zh}.

The following RHP uniquely determines $a_l^2(T)$ and partition function of the YM theory with the GWW potential.
Let $\Sigma$ be a unit circle with the counter clock-wise orientation and $Y(z,l+1,T)$ is a $2\times 2$ matrix-valued function satisfying
\begin{equation}
\begin{cases}
Y(z,l+1,T)\ \textit{is analytic in}\ \mathds{C}-\Sigma,\\
Y_+(z,l+1,T)=Y_-(z,l+1,T)\left(\begin{array}{cc}
1&\frac{1}{z^{l+1}}e^{\frac{T}{2}(z+z^{-1})}\\
0&1
\end{array}\right)\\
Y(z,l+1,T)z^{-(l+1)\sigma_3}=I+\mathcal{O}(\frac{1}{z})\ \textit{as}\ z\rightarrow \infty
\end{cases},
\end{equation}
where $Y_{\pm}$ denotes $Y$ inside/outside of the unit circle, $\sigma_3=\left(\begin{array}{cc}
1&0\\
0&-1
\end{array}\right)$ and $z^{-(l+1)\sigma_3}=\left(\begin{array}{cc}
z^{-(l+1)}&0\\
0&z^{l+1}
\end{array}\right)$.
It has been proved that the largest coefficient in the orthogonal polynomial is given by $21-$entry of matrix $Y$,
\begin{equation}
a_l^2(T)=-Y_{21}(0;l+1, T).
\end{equation}
Asymptotic results for $a_N^2(T)$ and $Z_{YM}^{(GWW)}$ for $N,T\rightarrow \infty$ is obtained in \cite{Ba-De-Jo}.
\section{Tracy-Widom distribution}\label{App: AppendixB}
Tracy-Widom distribution $F (x)$, \cite{Tr-Wi}, \cite{Ba-De-Jo}, is the probability distribution of the largest eigenvalue of a Hermitian random matrix and it is defined by
\begin{equation}
F(x)=e^{-\int_{x}^{\infty}(s-x)q^2(s)ds},
\end{equation}
where $q (s)$ is the solution of the Painlev\'{e} II equation, $q''(s)=2q^3 (s)+sq (s)$, with boundary condition $q(s)\sim -Ai(s)$ as $s\rightarrow \infty$, where $Ai(s)$ is the Airy function,
\begin{equation}
  Ai(x)=\frac{1}{2\pi}\int_{-\infty}^\infty e^{\frac{\ii s^3}{3}+\ii x s} ds=\frac{1}{\pi}\int_{0}^\infty \cos(\frac{s^3}{3}+ x s) ds.
\end{equation}
The Tracy-Widom distribution is also given by $F(x)=\det(1+K_{Ai})$ where the kernel of Airy function is given by
\begin{equation}
K_{Ai}(z,w)=\frac{Ai (z)Ai'(w)-Ai (w)Ai'(z)}{z-w}.
\end{equation}

In the asymptotic regime, $s\rightarrow \pm \infty$, the following result is obtained in \cite{Ba-Bu-Di},
\begin{equation}
q(s)=
\begin{cases}
-Ai(s)+\mathcal{O}( \frac { e^{-\frac {4}{3}s^{\frac {3}{2}}} }{s^{\frac {1}{4}}})&\ s\rightarrow+\infty\\
\sqrt{\frac{-s}{2}}(1+\frac{1}{8s^3}-\frac{73}{128s^6}+\frac{1020}{1024s^9}+\mathcal{O}(|s|^{-12}))&\ s\rightarrow-\infty
\end{cases},
\end{equation}
where in the limit $s\rightarrow +\infty$, $Ai''(s)=sAi (s)$ and $Ai(s)\approx \frac { e^{-\frac {2}{3}s^{\frac {3}{2}}} }{2\sqrt {\pi}s^{\frac {1}{4}}}$.

Moreover, the asymptotic of the Tracy-Widom distribution is studied in \cite{Ba-Bu-Di}, and the following result is obtained,
\begin{equation}\label{eq: asymp of Tracy Widom}
F(x)=
\begin{cases}
1-\frac{e^{-\frac{4}{3}x^{\frac{3}{2}}}}{32\pi x^{\frac{3}{2}}}(1-\frac{35}{24x^{\frac{3}{2}}}+\mathcal{O}(x^{-3}))& \ \text{as}\ x\rightarrow \infty\\
2^{\frac{1}{42}}e^{(\zeta^*)'(-1)}\frac{e^{-\frac{1}{12}|x|^{3}}}{|x|^{\frac{1}{8}}}(1+\frac{3}{2^6|x|^3}+\mathcal{O}(|x|^{-6}))& \ \text{as}\ x\rightarrow -\infty
\end {cases},
\end{equation}
where $\zeta^*$ is the Riemann zeta function.
\end{appendices}
{}
\vspace{10pt}
\address{Department of Mathematics and Statistics, P.O. Box 68, FIN\textendash{}00014
University of Helsinki, Finland}\\
\email{\textit{E-mail address}: seyedali.zahabi@helsinki.fi}

\begin{thebibliography}{}
\bibitem[Ah]{Ah} O. Aharony, S. Giombi, G. Gur-Ari, J. Maldacena and R. Yacoby, The thermal free energy in large N Chern-Simons-matter theories, Journal of High Energy Physics, \textbf{2013.3}, 1-38, 2013.
\bibitem[Ah-Gu-Ya1]{Ah-Gu-Ya1} O. Aharony, G. Gur-Ari and R. Yacoby, $D= 3$ bosonic vector models coupled to Chern-Simons gauge theories, Journal of High Energy Physics, \textbf{2012.3}, 1-25, 2012.
\bibitem[Ah-Gu-Ya2]{Ah-Gu-Ya2} O. Aharony, G. Gur-Ari and R. Yacoby, Correlation functions of large N Chern-Simons-Matter theories and bosonization in three dimensions, Journal of High Energy Physics, \textbf{2012.12}, 1-37, 2012.
\bibitem[Ar]{Ar} X. Arsiwalla, R. Boels, M. Marino and A. Sinkovics, Phase transitions in q-deformed 2d Yang-Mills theory and topological strings, Physical Review \textbf{D73.2}, 026005, 2006.
\bibitem[Ba-Bu-Di]{Ba-Bu-Di} J. Baik, R. Buckingham and J. DiFranco, Asymptotics of Tracy-Widom distributions and the total integral of a Painlev\'{e} II function, Communications in Mathematical Physics, \textbf{280.2}, 463-497, 2008.
\bibitem[Ba-De-Jo]{Ba-De-Jo} J. Baik, P. Deift and K. Johansson, On the distribution of the length of the longest increasing subsequence of random permutations, Journal of the American Mathematical Society, \textbf{12.4}, 1119-1178, 1999.
\bibitem[Ba-Li]{Ba-Li} J. Baik and Z. Liu, Discrete Toeplitz/Hankel determinants and the width of nonintersecting processes, International Mathematics Research Notices, \textbf{143}, 2013.
\bibitem[Ba-He]{Ba-He} S. Banerjee, S. Hellerman, J. Maltz and S. H. Shenker, Light states in Chern-Simons theory coupled to fundamental matter, Journal of High Energy Physics, \textbf{2013.3}, 1-45, 2013.
\bibitem[Bl-Th]{Bl-Th} M. Blau, and G. Thompson, Derivation of the Verlinde formula from Chern-Simons theory and the G/G model, Nuclear Physics \textbf{B408.2}, 345-390, 1993.
\bibitem[Ca]{Ca} N. Caporaso, M. Cirafici, L. Griguolo, S. Pasquetti, D. Seminara and R. J. Szabo, Topological strings and large N phase transitions I: Nonchiral expansion of q-deformed Yang-Mills theory, Journal of High Energy Physics, \textbf{2006.01}, 035, 2006.
\bibitem[Ch-Jo]{Ch-Jo} Y. Chen, N. Jokela, M. J\"{a}rvinen and N. Mekareeya, Moduli space of supersymmetric QCD in the Veneziano limit, Journal of High Energy Physics, \textbf{2013.9}, 1-38, 2013.
\bibitem[Ch-Mi]{Ch-Mi} C.-M. Chang, S. Minwalla, T. Sharma and X. Yin, ABJ triality: from higher spin fields to strings, Journal of Physics A: Mathematical and Theoretical, \textbf{46.21}, 214009, 2013.
\bibitem[De-Zh]{De-Zh} P. Deift and X. Zhou, A steepest descent method for oscillatory Riemann-Hilbert problems Asymptotics for the MKdV equation, Annals of Mathematics,  \textbf{295-368}, 1993.
\bibitem[Do-Ka]{Do-Ka} M.R. Douglas and V.A. Kazakov, Large N phase transition in continuum QCD 2, Physics Letters \textbf{B319.1}, 219-230, 1993.
\bibitem[Gi]{Gi} S. Giombi, S. Minwalla, S. Prakash, S.P. Trivedi and S.R. Wadia, Chern-Simons theory with vector fermion matter, European Physical Journal C-Particles and Fields, \textbf{72.8}, 1, 2012.
\bibitem[Gu]{Gu} G. Gur-Ari and R. Yacoby, Correlators of large N fermionic Chern-Simons vector models, Journal of High Energy Physics, \textbf{2013.2}, 1-17, 2013.
\bibitem[Gr-Wi]{Gr-Wi} D.J. Gross and E. Witten, Possible third-order phase transition in the large-N lattice gauge theory, Physical Review D, \textbf{21.2}, 446, 1980.
\bibitem[Ja]{Ja} S. Jain, S. P. Trivedi, S. R. Wadia and S. Yokoyama, Supersymmetric Chern-Simons theories with vector matter, Journal of High Energy Physics, \textbf{2012.10}, 1-46, 2012.
\bibitem[Ja-Ma]{Ja-Ma} D. Jafferis and J. Marsano, A DK Phase Transition in q-Deformed Yang-Mills on $S^ 2$ and Topological Strings, arXiv: hep-th/0509004, 2005.
\bibitem[Jo-Ja-Ke]{Jo-Ja-Ke} N. Jokela, M. J\"{a}rvinen and E. Keski-Vakkuri, Electrostatics of Coulomb gas, lattice paths and discrete polynuclear growth, Journal of Physics A: Mathematical and Theoretical, \textbf{43.42}, 425006, 2010.
\bibitem[Ma-Zh1]{Ma-Zh1} J. Maldacena and A. Zhiboedov, Constraining conformal field theories with a higher spin symmetry, Journal of Physics A: Mathematical and Theoretical, \textbf{46.21}, 214011, 2013.
\bibitem[Ma-Zh2]{Ma-Zh2} J. Maldacena and A. Zhiboedov, Constraining conformal field theories with a slightly broken higher spin symmetry, Classical and Quantum Gravity, \textbf{30.10}, 104003, 2013.
\bibitem[Ma1]{Ma1} M. Marino, Chern-Simons theory, matrix integrals, and perturbative three-manifold invariants, Communications in Mathematical Physics, \textbf{253.1}, 25-49, 2005.
\bibitem[Ma2]{Ma2} M. Marino, Chern-Simons theory, matrix models, and topological strings, Vol. \textbf{131}, Oxford University Press, 2005.
\bibitem[Ma-Mc]{Ma-Mc} A. Martínez-Finkelshtein,  K. T.-R. McLaughlin and E. B. Saff, Szeg\"o orthogonal polynomials with respect to an analytic weight: canonical representation and strong asymptotics, Constructive approximation, \textbf{24.3}, 319-363, 2006.
\bibitem[Sc]{Sc} A. S. Schwarz, New topological invariants arising in the theory of quantized fields, Baku Topol. Conf. 1987.
\bibitem[Sh-Yi]{Sh-Yi} S. H. Shenker and X. Yin, Vector models in the singlet sector at finite temperature, arXiv:1109.3519, 2011.
\bibitem[Sz]{Sz} G. Szeg\"{o}, Orthogonal polynomials, Vol. 23, American Mathematical Soc., 4th Ed, New York, 1975.
\bibitem[Ta]{Ta} S. Jain, S. Minwalla, T. Sharma, T. Takimi, S.R. Wadia and S. Yokoyama, Phases of large $N$ vector Chern-Simons theories on $S^2\times S^1$, Journal of High Energy Physics, \textbf{2013.9}, 1-85, 2013.
\bibitem[Tak]{Tak} T. Takimi, Duality and higher temperature phases of large N Chern-Simons matter theories on $S^2\times S^1$, Journal of High Energy Physics, \textbf{2013.7}, 1-62, 2013.
\bibitem[Tr-Wi]{Tr-Wi} C. A. Tracy and H. Widom, On orthogonal and symplectic matrix ensembles, Communications in Mathematical Physics, \textbf{177.3}, 727-754, 1996.
\bibitem[Va]{Va} M. A. Vasiliev, Nonlinear equations for symmetric massless higher spin fields in (A) dS (d), Physics Letters B, \textbf{567.1} 139-151, 2003.
\bibitem[Wa]{Wa} S. R. Wadia, $N=\infty$ phase transition in a class of exactly soluble model lattice gauge theories, Physics Letters, \textbf{B93.4}, 403-410, 1980.
\bibitem[Wi]{Wi} E. Witten, Quantum field theory and the Jones polynomial, Communications in Mathematical Physics, \textbf{121.3}, 351-399, 1989.
\bibitem[Yo]{Yo} S. Yokoyama, Chern-Simons-fermion vector model with chemical potential, Journal of High Energy Physics, \textbf{2013.1}, 1-13, 2013.

\end{thebibliography}
\end{document}